\documentclass[10pt,twoside,twocolumn,english,aps,manuscript,superscriptaddress,prbt,aps, preprint,manuscript,showpacs]{revtex4}
\usepackage{lmodern}
\usepackage[T1]{fontenc}
\usepackage[latin9]{inputenc}
\usepackage{color}
\usepackage{babel}
\usepackage{array}
\usepackage{amsmath}
\usepackage{amssymb}
\usepackage{graphicx}
\usepackage[unicode=true,pdfusetitle,
 bookmarks=true,bookmarksnumbered=false,bookmarksopen=false,
 breaklinks=false,pdfborder={0 0 1},backref=false,colorlinks=false]
 {hyperref}

\makeatletter

\newcommand*\LyXThinSpace{\,\hspace{0pt}}
\newcommand{\lyxmathsym}[1]{\ifmmode\begingroup\def\b@ld{bold}
  \text{\ifx\math@version\b@ld\bfseries\fi#1}\endgroup\else#1\fi}

\providecommand{\tabularnewline}{\\}

\@ifundefined{textcolor}{}
{%
 \definecolor{BLACK}{gray}{0}
 \definecolor{WHITE}{gray}{1}
 \definecolor{RED}{rgb}{1,0,0}
 \definecolor{GREEN}{rgb}{0,1,0}
 \definecolor{BLUE}{rgb}{0,0,1}
 \definecolor{CYAN}{cmyk}{1,0,0,0}
 \definecolor{MAGENTA}{cmyk}{0,1,0,0}
 \definecolor{YELLOW}{cmyk}{0,0,1,0}
}

\makeatother

\begin{document}

\title{Ba$_{3}$M$_{x}$Ti$_{3-x}$O$_{9}$(M = Ir, Rh): A family of 5\textit{\textcolor{black}{d}}/4\textit{\textcolor{black}{d}}-based,
diluted quantum spin liquids}

\author{R. Kumar }

\affiliation{Department of Physics, Indian Institute of Technology Bombay, Powai,
Mumbai 400076, India}

\author{D. Sheptyakov}

\affiliation{Laboratory for Neutron Scattering and Imaging, Paul Scherrer Institut,
5232 Villigen-PSI, Switzerland}

\author{P. Khuntia}

\affiliation{Laboratoire de Physique des Solides Universite Paris-Sud, UMR CNRS91405
Orsay, France}

\affiliation{Max-Planck Institute for Chemical Physics of Solids, 01187 Dresden,
Germany}

\author{K. Rolfs }

\affiliation{Laboratory for Developments and Methods, Paul Scherrer Institut,
5232 Villigen PSI, Switzerland }

\author{P. G. Freeman}

\affiliation{Laboratory for Quantum Magnetism, Ecole Polytechnique Federale de
Lausanne (EPFL), CH 1015, Switzerland}

\affiliation{Jeremiah Horrocks Institute for Mathematics, Physics and Astronomy,
University of Central Lancashire, Preston, PR1 2HE United Kingdom}

\author{H. M. Rønnow}

\affiliation{Laboratory for Quantum Magnetism, Ecole Polytechnique Federale de
Lausanne (EPFL), CH 1015, Switzerland}

\author{Tusharkanti Dey}

\affiliation{Experimental Physics VI, Center for Electronic Correlations and Magnetism,
University of Augsburg, D-86159 Augsburg, Germany }

\affiliation{Department of Physics, Indian Institute of Technology Bombay, Powai,
Mumbai 400076, India}

\author{M. Baenitz }

\affiliation{Max-Planck Institute for Chemical Physics of Solids, 01187 Dresden,
Germany}

\author{A. V. Mahajan}

\email{mahajan@phy.iitb.ac.in}

\affiliation{Department of Physics, Indian Institute of Technology Bombay, Powai,
Mumbai 400076, India}

\date{\today}
\begin{abstract}
\textcolor{black}{We report the structural and magnetic properties
of the 4}\textit{\textcolor{black}{d }}\textcolor{black}{(M = Rh)
based and 5}\textit{\textcolor{black}{d }}\textcolor{black}{(M = Ir)
based systems Ba$_{3}$M$_{x}$Ti$_{3-x}$O$_{9}$ (nominally $x$
= 0.5, 1). The studied compositions were found to crystallize in a
hexagonal structure with the centrosymmetric space group }$P6_{3}/mmc$.\textcolor{black}{{}
The structures comprise of A$_{2}$O$_{9}$ polyhedra (with the A
site (possibly) statistically occupied by M and Ti) in which pairs
of transition metal ions are stacked along the crystallographic }\textit{\textcolor{black}{c}}\textcolor{black}{-axis.
These pairs form triangular bilayers in the $ab$-plane. The magnetic
Rh and Ir ions occupy these bilayers, diluted by Ti ions even for
$x$ = 1. These bilayers are separated by a triangular layer which
is dominantly occupied by Ti ions. From magnetization measurements
we infer strong antiferromagnetic couplings for all of the materials
but the absence of any spin-freezing or spin-ordering down to 2~K.
Further, specific heat measurements down to 0.35~K show no sign of
a phase transition for any of the compounds. Based on these thermodynamic
measurements we propose the emergence of a quantum spin liquid ground
state for Ba$_{3}$Rh$_{0.5}$Ti$_{2.5}$O$_{9}$, and Ba$_{3}$Ir$_{0.5}$Ti$_{2.5}$O$_{9}$,
in addition to the already reported Ba$_{3}$IrTi$_{2}$O$_{9}$. }
\end{abstract}


\maketitle

\section{introduction}

Transition metal oxides (TMOs) containing 4\textit{d} and 5\textit{d}
metal atoms are of continuing interest and are being explored extensively
\cite{1-Balents }. In particular, 5\textit{d-}based TMOs are believed
to host various exotic ground states because of a relatively weak
on-site Coloumb repulsion $U$ (compared to typical values for 3\textit{d}
TMOs), and a strong spin-orbit coupling (SOC) $\lambda$. Also, it
is now well recognized that an intricate competition among the three
energy scales (a) on-site Coulomb repulsion $(U)$, (b) crystal-field
splitting $(\Delta)$, and (c) SOC, opens up avenues for the existence
of a diverse spectrum of physical properties \cite{1-Balents ,2-Khaliuilin }.
Among such exotic ground states, quantum spin liquids (QSL) are being
widely discussed. \textcolor{black}{The quantum spin liquid state
is a highly entangled state of localized moments that does not break
any symmetry of the Hamiltonian down to $T$ = 0 K}\textbf{\textcolor{red}{{}
}}\cite{3-Balents QSL}\textbf{\textcolor{black}{.}} \textcolor{black}{The
experimental signatures of the QSL state have been suggested to be
(i) a Curie-Weiss susceptibility ($C/(T-\theta$)) with a negative
$\theta$, (ii) no indication of spin ordering or spin freezing such
as a peak in the temperature dependence of susceptibility or the presence
of a bifurcation in the field cooled and zero-field cooled susceptibility
below some temperature or perhaps a large ``frustration ratio''
$f=|\theta/T_{N}|$ where $T_{N}$ is the ordering temperature, (iii)
absence of a peak in the temperature dependence of heat capacity but
the presence of a magnetic contribution to the heat capacity with
a power law behaviour in temperature, etc. While 3}\textit{\textcolor{black}{d}}\textcolor{black}{-based
TMOs include several materials with }QSL \cite{3-Balents QSL} behavior,
very few examples of 4\textit{d}/5\textit{d} based QSLs have been
reported. Prior to the work of Dey \textit{et al}. on Ba$_{3}$IrTi$_{2}$O$_{9}$
\cite{4-Ba3IrTi2O9} and Ba$_{3}$YIr$_{2}$O$_{9}$ (HP phase) \cite{5-Ba3YIr2O9 HP},
to our knowledge, Na$_{4}$Ir$_{3}$O$_{8}$ \cite{6-Na4Ir3O8  }
was the only reported QSL among 5\textit{d}-based TMOs. However, it
appears that, no such dynamically disordered ground state manifold
has been realized so far in 4\textit{d}-based TMOs, where \textit{U}
and SOC are expected to be of comparable strengths. Ba$_{2}$YMoO$_{6}$
\cite{7- Ba2YMo-13  Valence Bond Glass on an fcc Lattice,8-Evidence for a collective spin-singlet ground state,9-Triplet and in-gap magnetic s quantum frustrated Ba2YMo,10-Ba2YMoO6-16-AC}
drew considerable attention, but it does not fulfill the criteria
of a QSL and subsequent measurements suggest this to be a valence
bond glass (VBG) rather than a QSL.

\textcolor{black}{The Ba$_{3}$MX$_{2}$O$_{9}$ series with M = Co,
Ni, Cu, Ru, Ir, and X= Sb, Ti, \cite{4-Ba3IrTi2O9,11-Ba3CoSb2O9,12-Ba3Ni,13- Ba3CuSb2O9,14-Ba3RuTi2O9}
displays a variety of ground states depending upon the electronic
states of the ions, strength of the SOC, and the crystal field energy.
However, Ba$_{3}$ZnIr$_{2}$O$_{9}$, where Ir is in the pentavalent
oxidation state, shows a spin-orbital liquid state because of the
comparable energy scales induced by the SOC and superexchange interactions
\cite{15-Ba3ZnIr2O9}. }Dey \textit{et al. }\cite{4-Ba3IrTi2O9} studied
a $J_{eff}=1/2$ two-dimensional (2D) Mott insulator Ba$_{3}$IrTi$_{2}$O$_{9}$
which has significant Ir/Ti site disorder resulting in an effective
dilution of the magnetic Ir$^{4+}$ planes. Nevertheless, it appears
to be an example of a spin-orbit driven spin liquid, which does not
order down to $0.35$~K inspite of a large antiferromagnetic Curie-Weiss
temperature $\theta_{\mathrm{CW}}$ ($\sim-$$107$~K).\textcolor{red}{{}
}\textcolor{black}{Very recently, }Ba$_{3}$IrTi$_{2}$O$_{9}$\textcolor{black}{{}
has also evinced theoretical interest and Becker }\textit{\textcolor{black}{et
al}}\textcolor{black}{. \cite{16-Simon} proposed it to be a model
example to realize the Heisenberg-Kitaev model on a triangular lattice.
}Catuneanu\textcolor{black}{{} }\textit{et al. }\cite{17- Ba3Ir Stripy}\textit{
}\textcolor{black}{predicted}\textit{ }\textcolor{black}{Ba$_{3}$IrTi$_{2}$O$_{9}$
to host a stripy ordered magnetic ground state. Many of the materials
in the Ba$_{3}$MX$_{2}$O$_{9}$ series including Ba$_{3}$CuSb$_{2}$O$_{9}$
\cite{18 Ba3Cu Nakasduji}, Ba$_{3}$IrTi$_{2}$O$_{9}$ \cite{4-Ba3IrTi2O9},
and Ba$_{3}$RuTi$_{2}$O$_{9}$ \cite{14-Ba3RuTi2O9} exhibit atomic
site disorder. According to a recent theoretical proposal of Smerald
}\textit{\textcolor{black}{et al.}}\textcolor{black}{{} \cite{19-MILA}
correlated lattice disorder along with delocalised orphan spins can
promote a spin-orbital liquid phase, and this was suggested to be
the case for Ba$_{3}$CuSb$_{2}$O$_{9}$ \cite{13- Ba3CuSb2O9,18 Ba3Cu Nakasduji,20-Ba3CuSb2O9  NMR}.
The recent theoretical interest to underline the basic physics behind
the family of materials with formula Ba$_{3}$MX$_{2}$O$_{9}$, (
M = Co, Ni, Cu, Ru, Ir, and X = Sb, Ti), further motivates us to explore
these materials in detail. }

\textcolor{black}{Our present study focusses on Ba$_{3}$M$_{x}$Ti$_{3-x}$O$_{9}$
(nominally $x$ = 0.5, 1), where M = Ir, Rh. The Ba$_{3}$IrTi$_{2}$O$_{9}$
system was previously reported to crystallize in the hexagonal space
group $P6_{3}mc$~(186). In this space group, Ir$^{4+}$ (5$d^{5}$,
$J_{eff}=1/2$) and Ti$^{4+}$ ($S=0$) occupy the 2b sites and form
structural dimers (along the $c$-direction) which then form triangular
bi-planes ($ab$-plane) with an ordered arrangement of Ir and Ti.
Between succesive bi-planes there is a non-magnetic triangular layer
of Ti-atoms at the 2a sites \cite{4-Ba3IrTi2O9}. Due to their similar
ionic radii, however, it is inevitable to have mixing among Ir$^{4+}$
and Ti$^{4+}$ site occupancies and Rietveld refinement carried out
on x-ray data suggested this intersite disorder. }

\begin{figure}
\begin{centering}
\includegraphics[clip,scale=0.07]{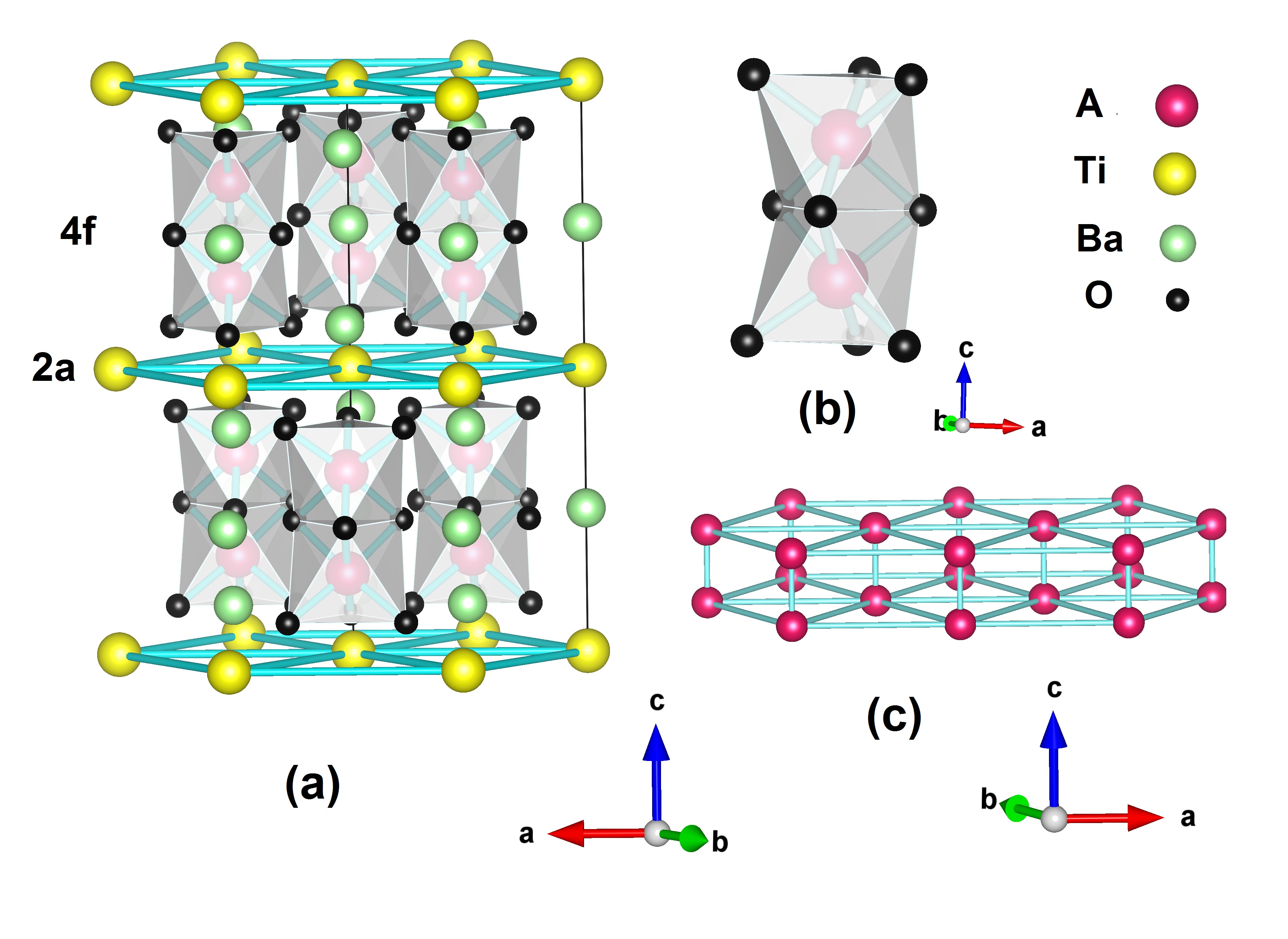}
\par\end{centering}

\protect\caption{\label{Fig 1}\textcolor{black}{{} }(Color online). Schematic representation
of the crystal structure of ${\rm Ba_{3}MTi_{2}O_{9}}$ (M=Ir,Rh)
in the \textcolor{black}{space group }$P6_{3}/mmc$. (a)~An ideal
unit cell i.e., with the $4f$ site equally occupied by Ir (or Rh)
and Ti and the $2a$ site occupied by Ti only; (b)~The structure
of a \textcolor{black}{A$_{2}$O$_{9}$}\ polyhedron (\textcolor{black}{where
A corresponds to M and Ti with a 50\% probability each}), consisting
of two face sharing ${\rm AO_{6}}$ octahedra, thus forming the M-Ti,
M-M, or Ti-Ti dimers; (c)~Formation of bi-triangular planes containing
an equal amount the nonmagnetic Ti and magnetic (Ir or Rh) atoms in
the $4f\,(1/3,2/3,z)$ positions.}
\end{figure}

\textcolor{black}{Here, we present neutron diffraction measurements
carried out on Ba$_{3}$IrTi$_{2}$O$_{9}$ and refinement with a
higher symmetry space group }$P6_{3}/mmc$~(the metal ions in the
dimers are disordered in this case)\textcolor{black}{{} which reveals
that approximately $56$\% of the 4f (1/3, 2/3, z) Wyckoff positions
are occupied by nonmagnetic Ti-atoms while the remaining $44$\% are
occupied by magnetic Ir-atoms. This gives rise to depleted bi-planes.
An idealized depiction (equal but disordered occupation of 4f sites
by M and Ti and sole occupation of 2a sites by Ti) of Ba$_{3}$MTi$_{2}$O$_{9}$(M
= Ir, Rh) in the space group }$P6_{3}/mmc$\textcolor{black}{{} can
be seen in Fig.~\ref{Fig 1}. }

\textcolor{black}{While Ba$_{3}$IrTi$_{2}$O$_{9}$, }Ba$_{3}$Ir$_{0.5}$Ti$_{2.5}$O$_{9}$,
and Ba$_{3}$Rh$_{0.5}$Ti$_{2.5}$O$_{9}$\textcolor{black}{{} could
be prepared in single phase form, Ba$_{3}$RhTi$_{2}$O$_{9}$ contained
large amounts of impurities. The large dilution which is naturally
present even for the nominal $x$ = 1 composition brings the system
in proximity of the percolation threshold (site percolation: p$_{c}$
= $0.5$) (see Ref. \cite{21- Percolation} ) of a 2D triangular lattice.}
For the presented materials, having triangular bilayers, the percolation
threshold might be somewhat smaller. In any case, in the systems studied
by us (single phase samples of nominal composition Ba$_{3}$IrTi$_{2}$O$_{9}$,
Ba$_{3}$Ir$_{0.5}$Ti$_{2.5}$O$_{9}$, and Ba$_{3}$Rh$_{0.5}$Ti$_{2.5}$O$_{9}$)
no ordering or freezing occurs down to $0.35$~K. The novelty of
these compounds is that even with such a large dilution (only about
25\% magnetic atoms in the bi-layers for the latter two), their magnetic
susceptibilities show a Curie-Weiss behaviour with a large Curie-Weiss
temperature (similar $\theta_{CW}$ for all the three systems), and
no ordering or freezing down to $0.35$~K. From our heat capacity
measurements we find that there is a magnetic contribution to the
heat capacity with a power-law variation with temperature. Our present
investigation suggests that the systems maintain a QSL state down
to $0.35$~K.\textbf{\textcolor{red}{{} }}

\section{experimental details }

Polycrystalline samples of nominal stoichiometry Ba$_{3}$IrTi$_{2}$O$_{9}$,
Ba$_{3}$Ir$_{0.5}$Ti$_{2.5}$O$_{9}$, Ba$_{3}$RhTi$_{2}$O$_{9}$,
and Ba$_{3}$Rh$_{0.5}$Ti$_{2.5}$O$_{9}$ were prepared by conventional
solid state reaction methods using high purity starting materials:
BaCO$_{3}$ (Alfa Aessar, 99.95\%), TiO$_{2}$ (Alfa Aessar, 99.9\%),
Ir (Arro Biochem, 99.9\%) and Rh (Alfa Aessar, 99.95\%) metal powder.
As will be detailed in the next section, for nominal compositions
of Ba$_{3}$IrTi$_{2}$O$_{9}$, Ba$_{3}$Ir$_{0.5}$Ti$_{2.5}$O$_{9}$,
Ba$_{3}$RhTi$_{2}$O$_{9}$ and Ba$_{3}$Rh$_{0.5}$Ti$_{2.5}$O$_{9}$,
the refined compositions were found to be \textcolor{black}{Ba$_{3}$Ir$_{0.946}$Ti$_{2.054}$O$_{9}$},
Ba$_{3}$Ir$_{0.484}$Ti$_{2.516}$O$_{9}$, ${\rm Ba_{3}Rh_{0.593}Ti_{2.407}O_{9}}$
and ${\rm Ba_{3}Rh_{0.432}Ti_{2.568}O_{9}}$, respectively.

Powder x-ray diffraction (xrd) measurements were performed at room
temperature with Cu $K_{\alpha}$ radiation ($\lambda=1.54182\,\lyxmathsym{\AA}$)
on a PANalytical X\textquoteright Pert PRO diffractometer. \textcolor{black}{Neutron
diffraction (ND) measurements were performed on the HRPT diffractometer
\cite{22-HRPT} at the SINQ neutron source of the Paul Scherrer Institut
(PSI) on the aforementioned polycrystalline samples at $300$~K and
$1.5$~K. In order to reduce the neutron absorption, the samples
(}Ba$_{3}$IrTi$_{2}$O$_{9}$\textcolor{black}{{} and }Ba$_{3}$RhTi$_{2}$O$_{9}$\textcolor{black}{)
were mounted in a vanadium sample can of $5.5$ mm diameter while
}Ba$_{3}$Ir$_{0.5}$Ti$_{2.5}$O$_{9}$\textcolor{black}{{} and }Ba$_{3}$Rh$_{0.5}$Ti$_{2.5}$O$_{9}$
\textcolor{black}{were mounted in a vanadium can of 7.5 mm diameter
with a sample height of approximately $30$~mm. Incident neutron
wavelengths of $1.15$}$\,\lyxmathsym{\AA}$\textcolor{black}{, $1.494$}$\,\lyxmathsym{\AA}$\textcolor{black}{,
and $1.88$}$\,\lyxmathsym{\AA}$\textcolor{black}{{} were selected
from a Bragg reflection from a Ge monochromator, with a pyrolytic
graphite filter placed in the incident beam to remove higher order
neutrons. The samples were mounted in a standard orange cryostat.
The scattered neutrons were detected over an angular range (2$\theta)$
}$3.75-164.7^{\circ}$\textcolor{black}{{} by a $^{3}$He multidetector.
A radial collimator was placed between the sample and detectors to
stop neutrons scattered from the sample environment from reaching
the detectors.} 

Magnetization measurements were carried out in the temperature range
$2-400$\,K and the field range $0-70$\,kOe using a Quantum Design
SVSM. Heat capacity measurements were performed in the temperature
range $0.35-295$\,K and in the field range $0-90$\,kOe using the
heat capacity option of a Quantum Design PPMS. 

\section{results}

In this section, we present the results of our systematic structural,
magnetic, and heat capacity measurements on the samples with nominal
compositions mentioned in the previous section.

\subsection{\textcolor{black}{Ba$_{3}$IrTi$_{2}$O$_{9}$} and Ba$_{3}$Ir$_{0.5}$Ti$_{2.5}$O$_{9}$ }

\subsubsection{Crystal structures }

\begin{figure*}[t]
\begin{centering}
\includegraphics[scale=0.6]{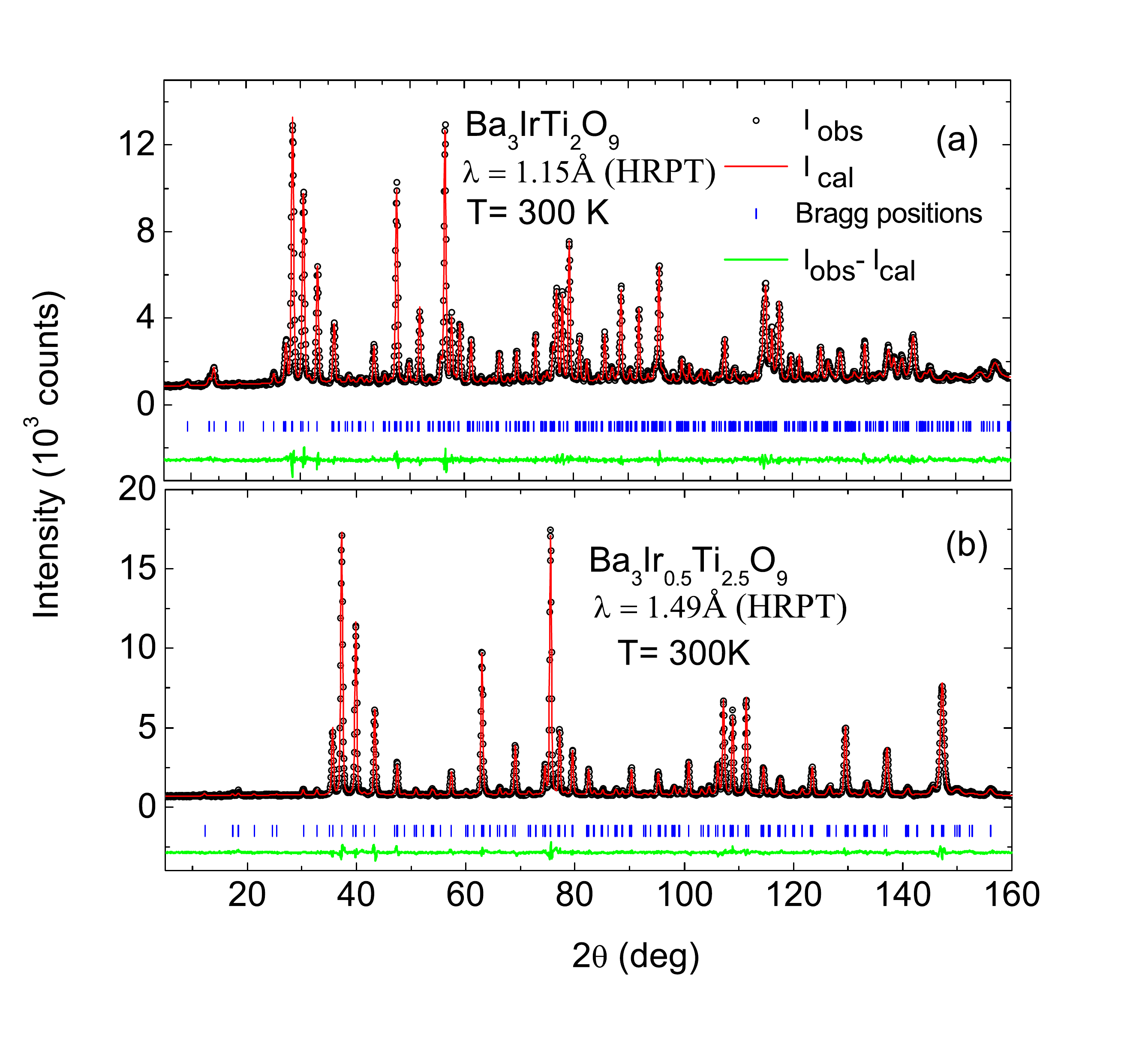}
\par\end{centering}

\protect\caption{(Color online). \label{Fig2-ND} Typical Rietveld refinement plots.
(a) The neutron diffraction data collected for Ba$_{3}$IrTi$_{2}$O$_{9}$
at $300$ K with wavelength $\lambda=1.15\textrm{\thinspace\AA}$
and (b) with wavelength $\lambda=1.49\textrm{\thinspace\AA}$ for
Ba$_{3}$Ir$_{0.5}$Ti$_{2.5}$O$_{9}$. Experimental points (black
open circles), calculated patterns (red lines) and difference curves
(green lines) are shown, the ticks (blue vertical lines) below the
graphs indicate the calculated positions of Bragg peaks. It should
be noted that the green lines on $y$-axis are shifted for clarity.}
\end{figure*}

The crystal structures of\textcolor{black}{{} Ba$_{3}$IrTi$_{2}$O$_{9}$}
and Ba$_{3}$Ir$_{0.5}$Ti$_{2.5}$O$_{9}$ have been investigated,
with particular emphasis on precise quantification of the metal site
occupation by Ir and Ti atoms. Both compounds display impurity-free
patterns and their refinements were carried out with the Fullprof
Suite of programs~\cite{23- FullProf}. \textcolor{black}{The typical
refinement plots for both the parent (Ba$_{3}$IrTi$_{2}$O$_{9}$)
and its magnetically diluted version (}Ba$_{3}$Ir$_{0.5}$Ti$_{2.5}$O$_{9}$\textcolor{black}{)
are presented in} Figs.~\textcolor{black}{\ref{Fig2-ND}(a)} and
(b), respectively. A strong contrast in scattering lengths between
Ir and Ti ($10.6$ and $-3.438\,fm$)~\cite{24- scattering length}
allows for a sufficiently high precision in determining the partial
occupations of Ir and Ti atoms at the $2a$ and $4f$ sites, respectively,
for the nominal $x$ = 1 composition. But, for the nominal $x$ =
0.5 composition, the small Ir~:~Ti ratio leads to a rather low average
scattering length, thus making the refinement of positional and atomic
displacement parameters less reliable or impossible. The crystal structure
of \textcolor{black}{Ba$_{3}$IrTi$_{2}$O$_{9}$} was reported by
Dey~\textit{et al.} (see Ref.~\cite{4-Ba3IrTi2O9}) to have the
space group $P6_{3}mc$ (No.~186, non-centrosymmetric). We have carried
out extensive refinements in both $P6_{3}mc$\ and $P6_{3}/mmc$\ (No.~194a,
centrosymmetric, higher symmetry supergroup of $P6_{3}mc$), and we
find that actually both choices lead to equally good refinements,
and essentially identical structures. The crystallographic sites occupied
by the Ti~:~Ir atoms in $P6_{3}/mmc$ space group model are: $2a\,(0,0,0)$
and $4f\,(1/3,2/3,z)$. In the non-centrosymmetric subgroup $P6_{3}mc$,
they transform into the $2a\,(0,0,z)$ (a site with floating $z$-coordinate),
and into a pair of $2b\,(1/3,2/3,z)$ sites correspondingly. Of significant
importance would be any statistically meaningful deviations in the
refined geometries and/or in the partial Ir~:~Ti occupations on
the metal sites, especially for the two individual metal positions
in the A$_{2}$O$_{9}$ double octahedral layers. However, from our
refinements in the lower symmetry $P6_{3}mc$ space group it follows
that for both \textcolor{black}{Ba$_{3}$IrTi$_{2}$O$_{9}$} and
Ba$_{3}$Ir$_{0.5}$Ti$_{2.5}$O$_{9}$ we get essentially identical
crystal structures as in the higher symmetry $P6_{3}/mmc$. 

\begin{table}
\begin{centering}
\protect\caption{\label{Ir and Ir0.5 table} Refined crystal structure parameters of
\textcolor{black}{Ba$_{3}$IrTi$_{2}$O$_{9}$}\ and Ba$_{3}$Ir$_{0.5}$Ti$_{2.5}$O$_{9}$.
Space group $P6_{3}/mmc$\ (No.~194), atoms are in the following
Wyckoff positions: Ba1 -- in $2b\,(0,0,1/4)$, Ba2 and (Ir/Ti)1 --
in $4f\,(1/3,2/3,z)$, (Ir/Ti)2 -- in $2a\,(0,0,0)$, O1 -- in $6h\,(x,2x,1/4)$,
and O2 -- in $12k\,(x,2x,z)$. Results obtained from neutron powder
data collected at 300~K with wavelengths $\lambda=1.1545\,\textrm{\AA}$
(\textcolor{black}{Ba$_{3}$IrTi$_{2}$O$_{9}$}) and $\lambda=1.494\,\textrm{\AA}$
(Ba$_{3}$Ir$_{0.5}$Ti$_{2.5}$O$_{9}$). Refined exact stoichiometries
of the two compounds are \textcolor{black}{Ba$_{3}$Ir$_{0.946}$Ti$_{2.054}$O$_{9}$}
and Ba$_{3}$Ir$_{0.484}$Ti$_{2.516}$O$_{9}$.}

\par\end{centering}

\begin{tabular}{lr>{\centering}p{3cm}>{\centering}p{2.8cm}}
\hline 
Atoms &  & \textcolor{black}{Ba$_{3}$IrTi$_{2}$O$_{9}$} & Ba$_{3}$Ir$_{0.5}$Ti$_{2.5}$O$_{9}$\tabularnewline
\hline 
 & $a,\textrm{\AA}$ & $5.7197(2)$ & $5.7239(1)$\tabularnewline
 & $c,\textrm{ \AA}$ & $14.0564(5)$ & $14.0070(2)$\tabularnewline
 & ${\rm V,\lyxmathsym{\AA}^{3}}$ & $398.25(3)$ & $397.43(1)$\tabularnewline
\hline 
Ba1 & $B_{iso},{\rm \lyxmathsym{\AA}^{2}}$ & $0.73(7)$ & $0.54(5)$\tabularnewline
Ba2 & $z$ & $0.5956(2)$ & $0.5971(2)$\tabularnewline
 & $B_{iso},{\rm \lyxmathsym{\AA}^{2}}$ & $0.73(4)$ & $0.69(3)$\tabularnewline
(Ir/Ti)1 & $z$ & $0.1606(3)$ & $0.1606$ \footnote{The $z$ coordinate for this position could not be refined because
of the unfavorable average scattering length for this combination
of Ti:Ir occupations (see text). It has therefore been fixed to the
value refined for \textcolor{black}{Ba$_{3}$IrTi$_{2}$O$_{9}$}.}\tabularnewline
 & $n\,(Ir\!:\!Ti)$ & $0.441(2)\!:\!0.559(2)$ & $0.242(3)\!:\!0.758(3)$\tabularnewline
 & $B_{iso},{\rm \lyxmathsym{\AA}^{2}}$ & $0.21(6)$ & $1.6$\footnote{for the same reason, this parameter was not refined, but instead fixed
to the value refined for the (Ti/Ir)2 position.}\tabularnewline
(Ir/Ti)2 & $n\,(Ir\!:\!Ti)$ & $0.064(4)\!:\!0.936(4)$ & $0.000(5)\!:\!1.000(5)$\tabularnewline
 & $B_{iso},{\rm \lyxmathsym{\AA}^{2}}$ & $0.88(12)$ & $1.6(1)$\tabularnewline
O1 & $x$ & $0.4854(5)$ & $0.4842(3)$\tabularnewline
 & $B_{iso},{\rm \lyxmathsym{\AA}^{2}}$ & $0.76(3)$ & $0.90(2)$\tabularnewline
O2 & $x$ & $0.1648(4)$ & $0.1655(3)$\tabularnewline
 & $z$ & $0.0795(1)$ & $0.0799(1)$\tabularnewline
 & $B_{iso},{\rm \lyxmathsym{\AA}^{2}}$ & $0.86(2)$ & $0.94(2)$\tabularnewline
 & $R_{p}$ & $2.91\%$ & $3.63\%$\tabularnewline
 & $R_{wp}$ & \textcolor{black}{{} }$3.68\%$ & $4.75\%$\tabularnewline
 & $R_{exp}$ & $2.44\%$ & $2.81\%$\tabularnewline
 & $\chi^{2}$ & $2.27$ & $2.86$\tabularnewline
\hline 
overall & $n\,(Ir\!:\!Ti)$ & $0.946(6)\!:\!2.054(6)$ & $0.484(8)\!:\!2.516(8)$\tabularnewline
refined &  &  & \tabularnewline
\hline 
\end{tabular}
\end{table}

This in particular implies that the partial Ir~:~Ti occupations
of both metal positions in the dimer layer of the face-sharing ${\rm MO_{6}}$
octahedra refine to the same values. Also, the refinement agreement
factors for the structure models for both space groups are practically
indistinguishable, which is why we consider it to be justified to
refine the crystal structure parameters of \textcolor{black}{Ba$_{3}$IrTi$_{2}$O$_{9}$}
and Ba$_{3}$Ir$_{0.5}$Ti$_{2.5}$O$_{9}$ in a model assuming the
higher symmetry space group $P6_{3}/mmc$~(No.~194). The refined
unit cell parameters at 300~K ($\{a,c\}=\{5.7197(2)\thinspace\textrm{\AA},14.0564(5)\,\textrm{\AA}\}$
and $\{5.7239(2)\,\textrm{\AA},14.0074(5)\,\textrm{\AA}\}$ for \textcolor{black}{Ba$_{3}$IrTi$_{2}$O$_{9}$}
and Ba$_{3}$Ir$_{0.5}$Ti$_{2.5}$O$_{9}$, correspondingly) are
in good agreement with the previously reported results~\cite{4-Ba3IrTi2O9}.
The refined crystal structure parameters are listed in Table~\textcolor{black}{\ref{Ir and Ir0.5 table}.}
The main finding is that in both compounds, the metal sites $2a\,(0,0,0)$
are predominantly occupied by Ti atoms (refined occupation ratios
of Ir to Ti are $0.064(4)\!:\negmedspace0.936(4)$ for \textcolor{black}{Ba$_{3}$IrTi$_{2}$O$_{9}$}
and $0.000(5)\!:\!1.000(5)$ for Ba$_{3}$Ir$_{0.5}$Ti$_{2.5}$O$_{9}$).
In the double layers of the stacked face-sharing ${\rm AO_{6}}$ octahedra,
the $n\,(Ir\!:\!Ti)$ ratios are $0.441(2)\!:\!0.559(2)$ and $0.242(3)\!:\!0.758(3)$
for \textcolor{black}{Ba$_{3}$IrTi$_{2}$O$_{9}$} and Ba$_{3}$Ir$_{0.5}$Ti$_{2.5}$O$_{9}$,
correspondingly, thus providing the final refined stoichiometries
of the two compounds: ${\rm Ba_{3}Ir_{0.946}Ti_{2.054}O_{9}}$ and
${\rm Ba_{3}Ir_{0.484}Ti_{2.516}O_{9}}$. These results imply that
Ir concentration in the biplanes in the Ba$_{3}$Ir$_{0.5}$Ti$_{2.5}$O$_{9}$
compound is $24\,\%$, while in the \textcolor{black}{Ba$_{3}$IrTi$_{2}$O$_{9}$}
compound, the biplanes have $44\,\%$ iridium. \textcolor{black}{The
}$2a\,(0,0,0)$\textcolor{black}{{} Wyckoff position is occupied by
$6$\% Ir-atoms.}

\textcolor{black}{The obtained atomic coordinates and occupancies
refined for Ba$_{3}$IrTi$_{2}$O$_{9}$ with }$P6_{3}/mmc$\textcolor{black}{~space
group are listed in Table~\ref{Ir and Ir0.5 table}. The main difference
between the structure of Ba$_{3}$MTi$_{2}$O$_{9}$, as per the space
groups }$P6_{3}/mmc$\textcolor{black}{{} or }$P6_{3}mc$\textcolor{black}{{}
arises as a result of the arrangement of magnetic atoms in 4f and
2b Wyckoff positions, respectively. The 4f Wyckoff position in }$P6_{3}/mmc$\textcolor{black}{{}
has a disordered occupation by M and Ti atoms giving rise to A$_{2}$O$_{9}$
polyhedra composed of two face sharing AO$_{6}$, (see Fig.~\ref{Fig 1}(b)),
where A corresponds to M and Ti with a 50\% probability each. The
A$_{2}$O$_{9}$ polyhedra make up }\textbf{\textcolor{black}{triangular
bi-planes}}\textcolor{black}{{} (see Fig.~\ref{Fig 1}(c)) which are
$50$\% diluted. In contrast, the 2b Wyckoff positions in }$P6_{3}mc$\textcolor{black}{{}
have Ir making undiluted }\textbf{\textcolor{black}{single triangular
planes}}\textcolor{black}{{} (}\textit{\textcolor{black}{ab}}\textcolor{black}{-plane)
in the ideal case. }

\subsubsection{Magnetic susceptibility}

\begin{figure}
\begin{centering}
\includegraphics[scale=0.35]{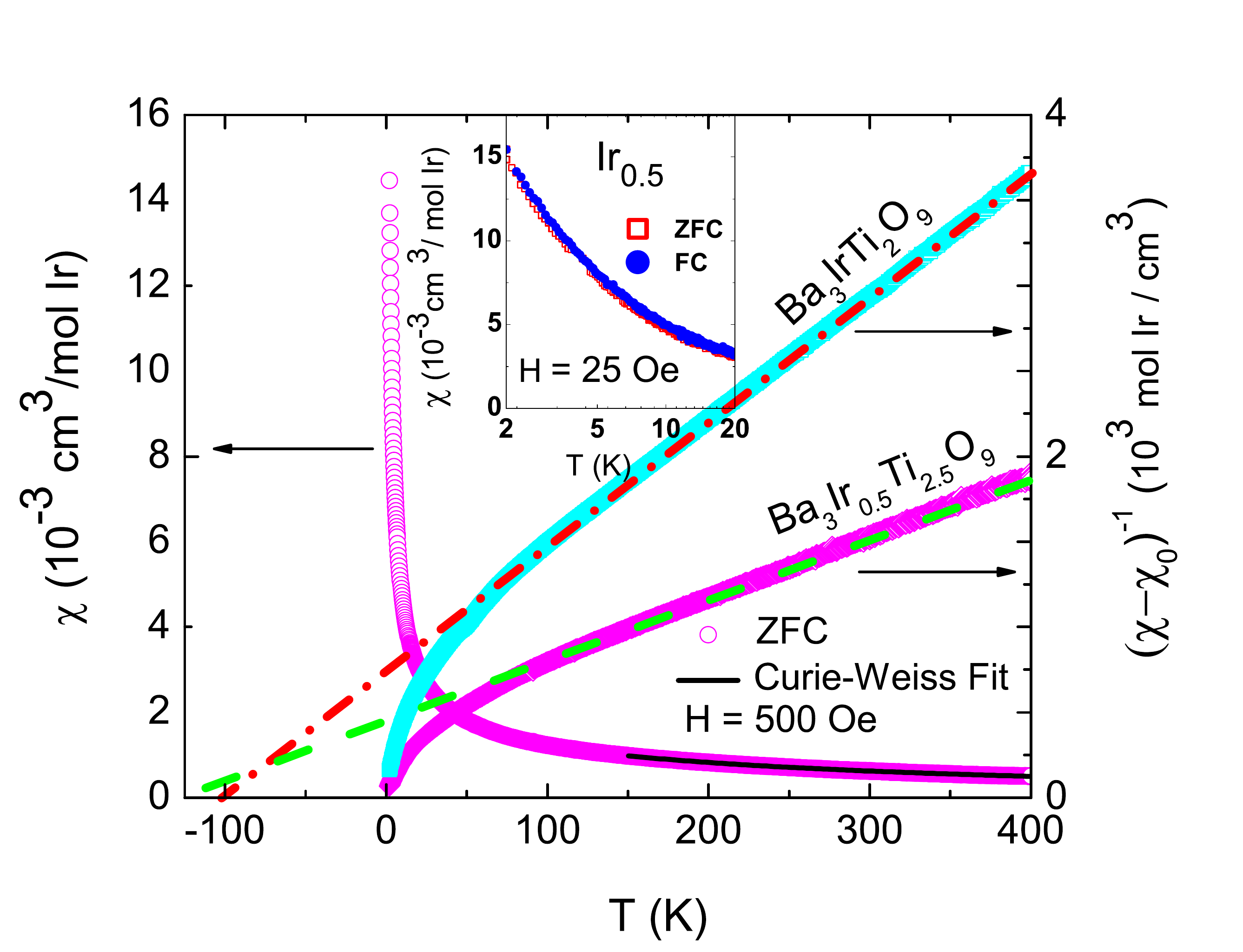}
\par\end{centering}

\protect\caption{(Color online). \label{Fig 3- Ir0.5 SVSM} The left \textit{y}-axis
shows the temperature dependent susceptibility of nominal Ba$_{3}$Ir$_{0.5}$Ti$_{2.5}$O$_{9}$
measured in $500$~Oe in the temperature range $2-400$~K and the
black solid line represents the Curie-Weiss fit in the temperature
range $150-400$~K. The  right \textit{y}-axis depicts the $T$-dependence
of inverse dc susceptibility of the parent compound Ba$_{3}$IrTi$_{2}$O$_{9}$(nominal)
and its diluted variant Ba$_{3}$Ir$_{0.5}$Ti$_{2.5}$O$_{9}$. Inset
shows the ZFC and FC data recorded in the temperature range $2-20$~K
at $25$~Oe for Ba$_{3}$Ir$_{0.5}$Ti$_{2.5}$O$_{9}$ in a linear-log
plot. }
\end{figure}

Magnetic susceptibility (as also its inverse) for nominal Ba$_{3}$Ir$_{0.5}$Ti$_{2.5}$O$_{9}$
is shown in Fig.~\ref{Fig 3- Ir0.5 SVSM}\textcolor{black}{.} For
clarity, only the inverse susceptibility variation with temperature
is shown for nominal Ba$_{3}$IrTi$_{2}$O$_{9}$. As illustrated
in the inset of Fig.~\ref{Fig 3- Ir0.5 SVSM}, the magnetization
measurements on Ba$_{3}$Ir$_{0.5}$Ti$_{2.5}$O$_{9}$ in a $25$~Oe
field do not evidence any bifurcation of the zero field cooled (ZFC)
and field cooled (FC) data. In contrast, a clear opening was observed
at $80$~K for the parent compound Ba$_{3}$IrTi$_{2}$O$_{9}$ which,
however closed (see supplemental information of Ref. \cite{4-Ba3IrTi2O9})
on increasing the field to $500$~Oe \cite{4-Ba3IrTi2O9}. The susceptibility
fit for Ba$_{3}$Ir$_{0.5}$Ti$_{2.5}$O$_{9}$ with the Curie-Weiss
law, $\chi=\chi_{0}+C/(T-\theta_{CW})$, in the temperature range
$150-400$~K yields a negative value of the asymptotic Curie-Weiss
temperature $\theta_{CW}\sim-125$~K and a temperature independent
susceptibility $\chi_{0}=-2.873\times10^{-5}$~cm$^{3}$/mole Ir.
Furthermore, a reduced value of the effective magnetic moment $\mu_{eff}=$
$1.32\pm0.05$$\,\mu_{B}$ was determined compared to the expected
1.73$\,\mu_{B}$ for a $S=1/2$ moment.  In fact, the parent, undiluted
compound shows an even bigger quenching of the magnetic moment ($\mu_{eff}=1.05\pm0.04\,\mu_{B}$
and $\theta_{CW}\sim-107$~K). A reduced $\mu_{eff}$ appears to
be common among iridates and might be a consequence of SOC. It is
not clear as to why a $50$\% magnetic dilution helps in increasing
the moment value.

\subsubsection{Heat capacity}

\begin{figure}[t]
\begin{centering}
\includegraphics[scale=0.355]{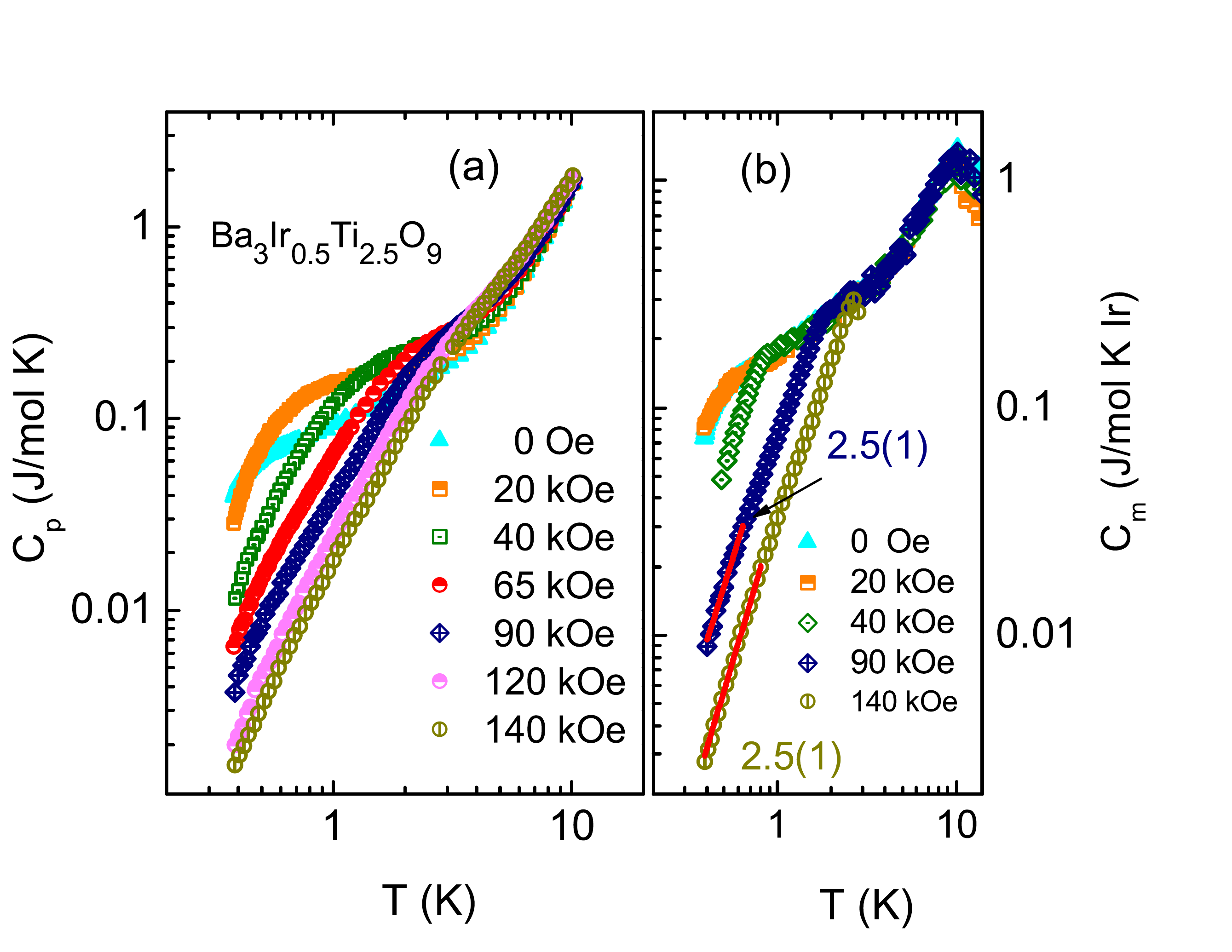}
\par\end{centering}

\protect\caption{(Color online). \label{Ir0.5 Cp(T)} (a) The $C_{p}(T)$ data for
Ba$_{3}$Ir$_{0.5}$Ti$_{2.5}$O$_{9}$ (baked) is depicted in the
log-log plot as a function of temperature at different fields. (b)
The magnetic heat capacity $C_{m}$ is shown in a log-log plot and
the solid lines are fits to the power law: $C_{m}=\gamma T^{\alpha}$
(explained in the text). The obtained exponent values are written
against the $C_{m}$ data.}
 
\end{figure}

To explore further the ground state properties of nominal Ba$_{3}$Ir$_{0.5}$Ti$_{2.5}$O$_{9}$,
we examine heat capacity $C_{P}(T)$ data in various fields ($0-140$~kOe),
as shown in Fig.~\ref{Ir0.5 Cp(T)}(a). No anomaly, hallmark of any
short-range or long-range magnetic ordering, was observed in the $C_{P}(T)$
data in the measured temperature range. However, below about $10$~K,
$C_{P}(T)$ data show a broad maximum which shifts towards higher-$T$
upon increasing the magnetic field indicating that it might be a Schottky
contribution originating from orphan spins as is apparently very common
for disordered materials. These orphan spins can arise from (a) Ir
occupying the $2a$ sites and (b) a small fraction of uncompensated
Ir$^{4+}$ spins from the triangular bilayers. Such orphan spins then
might form a two-level system and in turn show a field dependence
at low-$T$. This behavior was also noticed in the parent compound
Ba$_{3}$IrTi$_{2}$O$_{9}$~(nominal) \cite{4-Ba3IrTi2O9}.\textcolor{red}{{} }

\textcolor{black}{The specific heat of a magnetic insulator in the
presence of orphan spins comprises of (a) specific heat due to the
lattice ($C_{lattice}$), (b) specific heat due to uncorrelated spins
or the Schottky specific heat ($C_{Sch}$), and (c) specific heat
due to correlated magnetic spins ($C_{m}$). In the absence of a suitable
nonmagnetic analog, $C_{lattice}$ for Ba$_{3}$Ir$_{0.5}$Ti$_{2.5}$O$_{9}$
was estimated by adopting Ba$_{3}$ZnSb$_{2}$O$_{9}$ as the non-magnetic
analog. The $C{}_{P}(T)$ data for Ba$_{3}$ZnSb$_{2}$O$_{9}$ were
extracted from Ref.~\cite{13- Ba3CuSb2O9}. The heat capacity of
this nonmagnetic analog was scaled to match with the heat capacity
of Ba$_{3}$Ir$_{0.5}$Ti$_{2.5}$O$_{9}$ in the temperature range
$20-45$~K. The corrected heat capacity of Ba$_{3}$ZnSb$_{2}$O$_{9}$
thus obtained was subtracted from the measured heat capacity of Ba$_{3}$Ir$_{0.5}$Ti$_{2.5}$O$_{9}$.
This yielded $C_{Sch}$ + $C_{m}$. }

\begin{figure}[h]
\centering{}\includegraphics[scale=0.35]{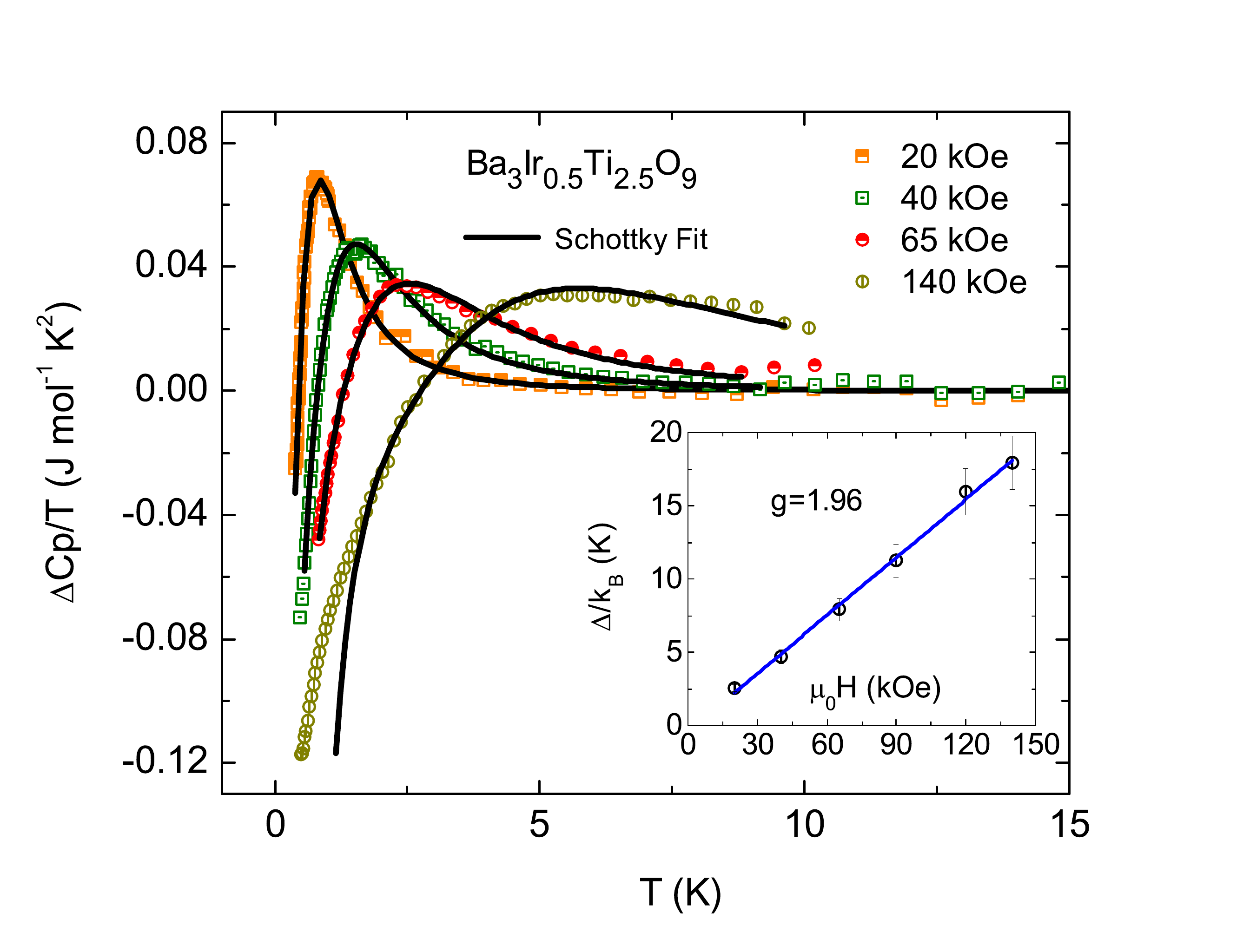}\protect\caption{(Color online). \label{Schottky gap Ir0.5} A few representative plots
of $[C_{P}(H)-C_{P}(H=0)]/T$ or ($\triangle C_{P}/T$) for Ba$_{3}$Ir$_{0.5}$Ti$_{2.5}$O$_{9}$
at various values of the field $H$ (as given in the legend) and their
fit to Eq.~\ref{eq:1} (black lines). Inset: the variation of the
Schottky gap as a function of magnetic field for orphan spins. }
\end{figure}

To remove the Schottky contribution from $C_{P}(T)$ data, the following
strategy was adopted: first we subtracted the $C_{P}(H=0)$ data from
those measured in $H$ $\neq0$, i.e., $C_{P}(H)$. This difference
($\triangle C_{P}/T$) was modeled as a function of temperature using
$[C_{Sch}(\triangle_{1})-C_{Sch}(\triangle_{2})]f/T$, where $f$
stands for the percentage of orphan spins present in the sample and
$C_{Sch}(\triangle_{1})$ and $C_{Sch}(\triangle_{2})$ are the Schottky
contributions to the specific heat. Figure \ref{Schottky gap Ir0.5}
shows the estimated contribution as well as its variation (using Eq.~\ref{eq:1})
as a function of temperature for several magnetic fields. The parameters
$\triangle_{1}$ and $\triangle_{2}$ are the values of energy separation
under the influence of external magnetic fields $H_{1}$ and $H_{2}$(=
$0$~ Oe), respectively. Furthermore, $C_{Sch}$ is defined by the
equation: 

\begin{center}
\begin{equation}
C_{Sch}(\triangle)=R(\frac{\triangle}{k_{B}T})^{2}(\frac{g_{0}}{g_{1}})\frac{exp(\frac{\triangle}{k_{B}T})}{[1+(\frac{g_{0}}{g_{1}})exp(\frac{\triangle}{k_{B}T})]^{2}}\label{eq:1}
\end{equation}
 
\par\end{center}

Here, $R$ and $k{}_{B}$ are the universal gas constant and the Boltzmann
constant, respectively. The parameters $g{}_{0}$ and $g{}_{1}$ are
the degeneracies of the two level system and in the present case are
assumed to be unity. This routine process to eliminate the orphan
spin contribution has already been used earlier in materials such
as Ba$_{3}$CuSb$_{2}$O$_{9}$ \cite{13- Ba3CuSb2O9} and ZnCu$_{3}$(OH)$_{6}$Cl$_{2}$
\cite{25-Kagome Schottky}. While estimating the Schottky contribution,
the parameters $\triangle_{1}$, $\triangle_{2}$ and $f$ were kept
free and the\textcolor{black}{{} variation of} $\triangle_{1}$\textcolor{black}{{}
for various magnetic fields is shown in the inset of Fig. \ref{Schottky gap Ir0.5}.
The parameter }$\triangle_{2}$\textcolor{black}{{} was found to vary
between $0.3$~K and $1.44$~K (not significantly different from
zero) in the field range $20-140$~kOe.} The extracted value of $f$
for Ba$_{3}$Ir$_{0.5}$Ti$_{2.5}$O$_{9}$ amounts to $3$\%-$4$\%
and might come from Ir$^{4+}$ spins. Estimated $g$-value (see inset
of Fig. \textcolor{black}{\ref{Schottky gap Ir0.5}} ) obtained for
orphan spins is $1.96$, which is the free spin only value suggesting
that orphan spins might not be affected by SOC. 

Finally the inferred $C_{m}$ is plotted in Fig.~\ref{Ir0.5 Cp(T)}(b).
In an attempt to shed some light on the nature of magnetic excitations
at low-$T$, the $C_{m}$ data were fitted with the equation $C_{m}$
=$\gamma$$T$$^{\alpha}$. The obtained value of $\alpha$ for\textcolor{black}{{}
$T\leq$$1$~K} varies somewhat with $H$ but is nearly constant
at about $2.5$ in high fields. Such $H$-dependent behavior of $C_{m}$
was also observed for compound Ba$_{3}$IrTi$_{2}$O$_{9}$ \cite{4-Ba3IrTi2O9}.
\textcolor{black}{An uncertainty in determining$C_{Sch}$ at lower
fields, where the Schottky hump lies at low-$T$ might affect the
value of }$\alpha$\textcolor{black}{. But for $H$ $\geqslant90$~kOe,
where the Schottky hump shifts to higher-$T$, the obtained }$\alpha$\textcolor{black}{{}
for $T\leq1~$K remains unaffected by $C_{Sch}$ because of its vanishingly
small contribution to }$C_{P}(T)$\textcolor{black}{.}

The magnetic entropy change $\triangle S_{m}$ was calculated by integrating
$\frac{C_{m}}{T}$ vs $T$ data for the Ir$_{0.5}$ sample, as shown
in Fig.~\ref{Entropy change}. The calculated $\triangle S_{m}$
in zero field was found to be $1.76$ (J/K mol Ir) below $25$~K,
which is nearly $30$\% of $Rln2=5.76$ J/mol K, expected for a $S=1/2$
system. In contrast, for Ba$_{3}$IrTi$_{2}$O$_{9}$, it is $10$\%
of what is expected for a $S=1/2$ system. This suggests that the
diluted compound is less magnetically disordered but the higher effective
moment in the Ir$_{0.5}$ system may also play a role.

\begin{figure*}
\begin{centering}
\includegraphics[scale=0.6]{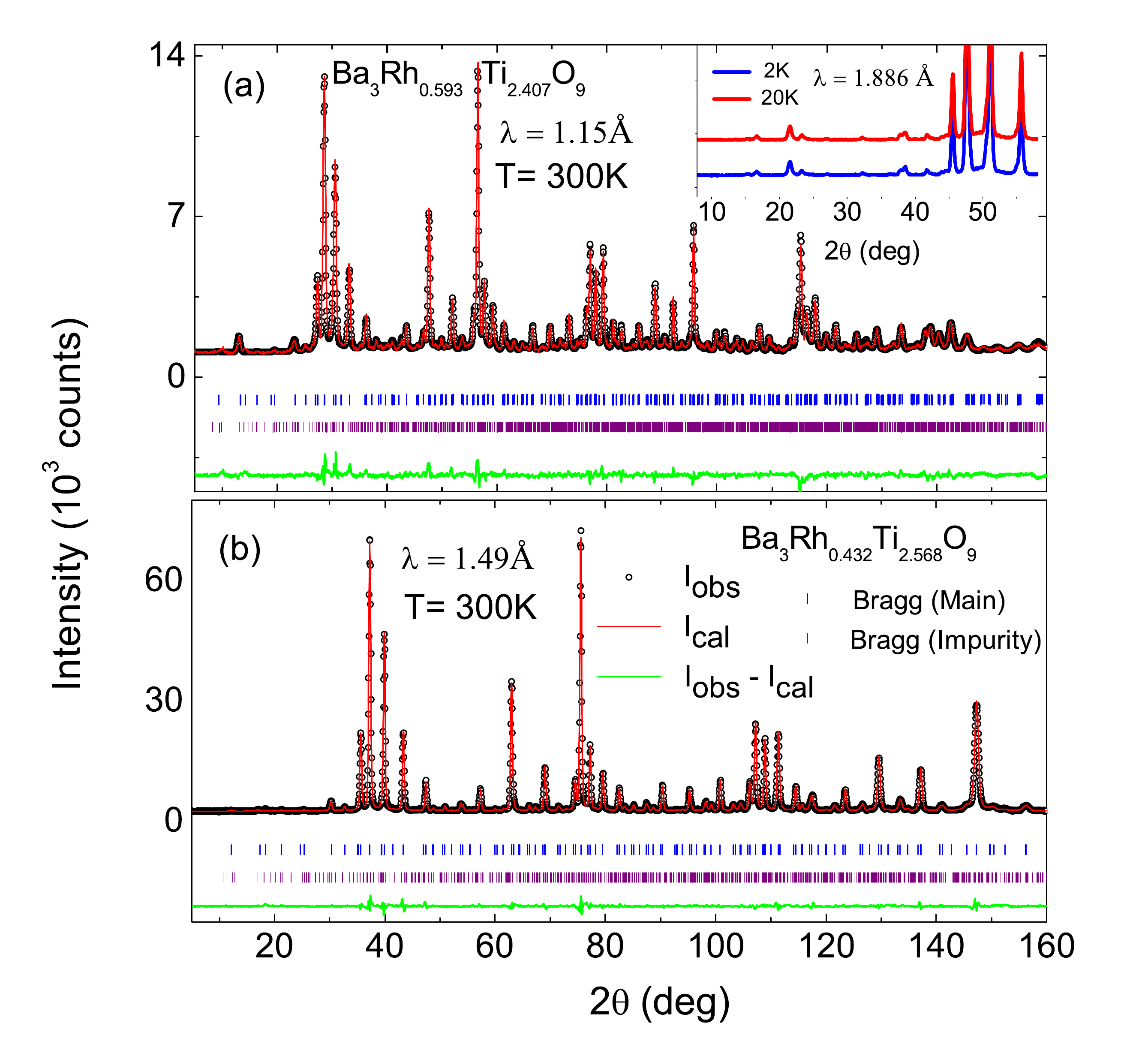}
\par\end{centering}

\protect\caption{(Color online). \label{Rh=000026Rh0.5 ND} The neutron diffraction
data collected at 300~K for \textcolor{black}{(a) }nominal\textcolor{black}{{}
Ba$_{3}$RhTi$_{2}$O$_{9}$ (in actuality }${\rm Ba_{3}Rh_{0.593}Ti_{2.407}O_{9}}$\textcolor{black}{)
with }$\lambda=1.15\,\textrm{\AA}$\textcolor{black}{{} and (b) }nominal\textcolor{black}{{}
Ba$_{3}$Rh$_{0.5}$Ti$_{2.5}$O$_{9}$ (in actuality} ${\rm Ba_{3}Rh_{0.432}Ti_{2.568}O_{9}}$\textcolor{black}{)
with }$\lambda=1.49\,\textrm{\AA}$,\textcolor{black}{{} are shown with
open black circles and the red solid lines represent the Rietveld
refined patterns. The Bragg positions for main and impurity phases
(}Ba$_{9}$Rh$_{8}$O$_{24}$\textcolor{black}{) are shown by vertical
blue and purple lines and the difference patterns are depicted} with
green lines, respectively. It should be noted that Ba$_{3}$RhTi$_{2}$O$_{9}$(nominal)
contains roughly $12$\% of impurity phase Ba$_{9}$Rh$_{8}$O$_{24}$,
whereas its diluted variant Ba$_{3}$Rh$_{0.5}$Ti$_{2.5}$O$_{9}$
(nominal) possesses less than $1.5$\% of impurity phase of Ba$_{9}$Rh$_{8}$O$_{24}$.
The inset shows a comparison of the low-angle ranges of the experimental
neutron diffraction patterns of \textcolor{black}{Ba$_{3}$RhTi$_{2}$O$_{9}$}\ at
20 and at 2~K with $\lambda=1.886\,\textrm{\AA}$, indicating that
no additional (magnetic) intensity can be observed at the lowest achieved
temperature. }
\end{figure*}

\subsection{Ba$_{3}$RhTi$_{2}$O$_{9}$ and Ba$_{3}$Rh$_{0.5}$Ti$_{2.5}$O$_{9}$ }

\subsubsection{Crystal Strucure}

In the following section, we focus on the $4d$-based Ba$_{3}$Rh$_{x}$Ti$_{3-x}$O$_{9}$\ which
is isostructural to the $5d$-based Ba$_{3}$IrTi$_{2}$O$_{9}$.
The Ba$_{3}$RhTi$_{2}$O$_{9}$\ was first reported by Bryne \textit{et
al.} \cite{26- Ba3RhTi2O9} and its xrd pattern was claimed to be
similar to that of the hexagonal ${\rm BaTiO_{3}}$, but also containing
some secondary, unidentified phase(s). However no further detailed
information on the crystal structure of Ba$_{3}$RhTi$_{2}$O$_{9}$\ was
provided. Our laboratory x-ray diffraction pattern (not shown) of
the nominal Ba$_{3}$RhTi$_{2}$O$_{9}$ has shown the extra lines
as well, which were identified as coming from ${\rm Ba_{9}Rh_{8}O_{24}}$
impurity \cite{27- Impurity}. Even with this impurity introduced
into the refinement, there are still a few spurious lines in the pattern,
due to low amounts of other impurities. Neutron powder diffraction
is again of great advantage for studying the crystal structure of
Ba$_{3}$Rh$_{x}$Ti$_{3-x}$O$_{9}$, since the difference in neutron
scattering lengths between Rh and Ti ($5.88$ and $-3.438\,fm$)~\cite{24- scattering length}
allows for precise quantification of the partial occupations of Rh
and Ti in different metal sites. Neutron powder diffraction data for
the compound with the intended composition Ba$_{3}$RhTi$_{2}$O$_{9}$
have been collected at 2~K with the wavelengths $\lambda=1.155\,\textrm{\AA}$
and $\lambda=1.886\,\textrm{\AA}$, at 20~K with $\lambda=1.886\,\textrm{\AA}$,
and at 300~K with $\lambda=1.1545\,\textrm{\AA}$. The shorter wavelength
data were collected mainly to achieve the best precision in the crystal
structure refinements, while the purpose to collect longer wavelength
data at 2 and 20~K was to make sure that no magnetic ordering would
manifest itself in appearance of any additional diffraction intensity
at the base temperature. Coherent with the observation of the laboratory
x-ray measurement, we see the ${\rm Ba_{9}Rh_{8}O_{24}}$ impurity
($\sim12\,wt.\%$), and a few weak additional spurious lines at essentially
the same $d$-spacing values as in the laboratory x-ray data.\textcolor{black}{{}
}Sample preparation under various conditions failed to produce a single
phase sample of Ba$_{3}$RhTi$_{2}$O$_{9}$. Consequently, we \textbf{\textcolor{black}{do
not}} report the magnetic and heat capacity data for Ba$_{3}$RhTi$_{2}$O$_{9}$.
We note, just for completeness that, the refined composition of this
compound is ${\rm Ba_{3}Rh_{0.593(13)}Ti_{2.407(18)}O_{9}}$ (the
deficit of rhodium is due to the large amount of Rh-rich ${\rm Ba_{9}Rh_{8}O_{24}}$
impurity), and that we do not observe an appearance of any additional
Bragg intensities at low~Q values when cooling down from 20 to 2~K
and thus -- down to the precision of our experiment, we can also exclude
an emergence of LRO in ${\rm Ba_{3}Rh_{0.593}Ti_{2.407}O_{9}}$. An
example of a Rietveld refinement done on the powder neutron diffraction
data of this compound collected at the base temperature is given in
the Fig.~\ref{Rh=000026Rh0.5 ND}(a). The refined crystal structure
parameters at room temperature are given in Table~\ref{Rh and Rh0.5 position table}. 

\begin{table}[t]
\begin{centering}
\protect\caption{\label{Rh and Rh0.5 position table} Refined crystal structure parameters
of the two studied Ba$_{3}$Rh$_{x}$Ti$_{3-x}$O$_{9}$\ compounds
at room temperature. Space group $P6_{3}/mmc$~(No.~194), atoms
are in the following Wyckoff positions: Ba1 -- in $2b\,(0,0,1/4)$,
Ba2 and (Ti/Rh)1 -- in $4f\,(1/3,2/3,z)$, (Ti/Rh)2 -- in $2a\,(0,0,0)$,
O1 -- in $6h\,(x,2x,1/4)$, and O2 -- in $12k\,(x,2x,z)$. Results
obtained from neutron powder data collected at 300~K with wavelengths
$\lambda=1.1545\,\textrm{\AA}$ (sample with target composition Ba$_{3}$RhTi$_{2}$O$_{9}$,
left column) and $\lambda=1.494\,\textrm{\AA}$ (sample with target
composition Ba$_{3}$Rh$_{0.5}$Ti$_{2.5}$O$_{9}$, right column).
Refined exact stoichiometries of the two compounds are ${\rm Ba_{3}Rh_{0.593}Ti_{2.407}O_{9}}$
and ${\rm Ba_{3}Rh_{0.432}Ti_{2.568}O_{9}}$.}

\par\end{centering}

\centering{}%
\begin{tabular}{lrcc}
\hline 
\multicolumn{2}{l}{Target composition} & \textcolor{black}{Ba$_{3}$RhTi$_{2}$O$_{9}$} & \textcolor{black}{Ba$_{3}$Rh$_{0.5}$Ti$_{2.5}$O$_{9}$}\tabularnewline
\hline 
\multicolumn{2}{l}{refined $(Rh\!:\!Ti)$} & $0.593(13)\!:\!2.407(13)$ & $0.432(18)\!:\!2.568(18)$\tabularnewline
composition  &  &  & \tabularnewline
\multicolumn{2}{l}{${\rm Ba_{9}Rh_{8}O_{24}}$} & 12(1) & 1.5(0.5)\tabularnewline
\multicolumn{2}{l}{impurity (wt.~\%)} &  & \tabularnewline
\hline 
 & $a,\textrm{\AA}$ & $5.7186(7)$  & $5.7209(1)$\tabularnewline
 & $c,\textrm{ \AA}$ & $14.0271(18)$  & $14.0159(2)$\tabularnewline
 & ${\rm V,\lyxmathsym{\AA}^{3}}$ & $397.26(9)$  & $397.27(1)$\tabularnewline
\hline 
Ba1  & $B_{iso},{\rm \lyxmathsym{\AA}^{2}}$ & $0.56(11)$$^{a}$ & $0.49(2)$$^{a}$\tabularnewline
Ba2  & $z$  & $0.5947(8)$ & $0.5952(2)$\tabularnewline
 & $B_{iso},{\rm \lyxmathsym{\AA}^{2}}$ & $0.56(11)$$^{a}$ & $0.49(2)$\footnote{The B$_{iso}$ parameters for both Ba atoms were refined with constraint
to equality}\tabularnewline
(Rh/Ti)1  & $z$  & $0.143(6)$  & $0.1480(5)$\tabularnewline
 & $n\,(Rh\!:\!Ti)$  & $0.292(5)\!:\!0.708(5)$  & $0.212(7)\!:\!0.788(7)$\tabularnewline
 & $B_{iso},{\rm \lyxmathsym{\AA}^{2}}$ & $1.0(2)$$^{b}$ & $0.57(6)$\tabularnewline
(Rh/Ti)2  & $n\,(Rh\!:\!Ti)$  & $0.009(8)\!:\!0.991(8)$  & $0.008(5)\!:\!0.992(5)$\tabularnewline
 & $B_{iso},{\rm \lyxmathsym{\AA}^{2}}$ & $1.0(2)$$^{b}$ & $0.57(6)$\footnote{The B$_{iso}$ parameters for both Ti:Ir atoms were refined with constraint
to equality}\tabularnewline
O1  & $x$  & $0.4840(17)$  & $0.4839(3)$\tabularnewline
 & $B_{iso},{\rm \lyxmathsym{\AA}^{2}}$ & $0.69(9)$  & $0.74(2)$\tabularnewline
O2  & $x$  & $0.1653(14)$  & $0.1648(2)$\tabularnewline
 & $z$  & $0.0796(3)$  & $0.0802(1)$\tabularnewline
 & $B_{iso},{\rm \lyxmathsym{\AA}^{2}}$ & $0.62(5)$  & $0.67(1)$\tabularnewline
 & $R_{p}\,(\%)$ & $3.21$ & $2.95$\tabularnewline
 & $R_{wp}\,(\%)$ & $4.02$ & $3.98$\tabularnewline
 & $R_{exp}\,(\%)$ & $2.41$ & $1.46$\tabularnewline
 & $\chi^{2}$ & $2.78$ & $7.40$\tabularnewline
\hline 
\end{tabular}
\end{table}

The second compound, with a target nominal composition Ba$_{3}$Rh$_{0.5}$Ti$_{2.5}$O$_{9}$,
did show a much smaller amount (< 1.5\%) of ${\rm Ba_{9}Rh_{8}O_{24}}$
impurity \cite{27- Impurity}, and the refined composition is ${\rm Ba_{3}Rh_{0.432(18)}Ti_{2.568(18)}O_{9}}$.
This sample is then single phase. The Rietveld refined pattern is
shown in Fig.~\ref{Rh=000026Rh0.5 ND}(b). Practically all crystal
structure parameters (given in the right column in Table~\ref{Rh and Rh0.5 position table})
are very close to those found for ${\rm Ba_{3}Rh_{0.593}Ti_{2.407}O_{9}}$.
Thus, in our synthetic approaches, the $\sim0.5:2.5$ ratio between
Rh and Ti amounts in the Ba$_{3}$Rh$_{1-x}$Ti$_{2+x}$O$_{9}$\ structure
seems to be a characteristic feature, the remaining Rh being dumped
into the Rh-rich impurity (in our case ${\rm Ba_{9}Rh_{8}O_{24}}$). 

From our results it follows that in both studied Ba$_{3}$Rh$_{x}$Ti$_{3-x}$O$_{9}$
compositions, the metal $2a(0,0,0)$ sites are practically exclusively
occupied by titanium, while the Rh~:~Ti occupation factors ratio
of roughly 1~:~3 is characteristic for the $4f(1/3,2/3,z)$ sites.
This delivers the refined compositions ${\rm Ba_{3}Rh_{0.593}Ti_{2.407}O_{9}}$
and ${\rm Ba_{3}Rh_{0.432}Ti_{2.568}O_{9}}$ for the two studied compounds,
thus indicating a stability (under our synthetic conditions) of the
approximate Ba$_{3}$Rh$_{0.5}$Ti$_{2.5}$O$_{9}$ composition. Since
the nominal Ba$_{3}$RhTi$_{2}$O$_{9}$ is not single phase, we report
further measurements on only Ba$_{3}$Rh$_{0.5}$Ti$_{2.5}$O$_{9}$
which is single phase. This strongly diluted system seems to show
no sign of LRO down to 2~K.

\subsubsection{Magnetic susceptibility}

\begin{figure}
\begin{centering}
\includegraphics[scale=0.37]{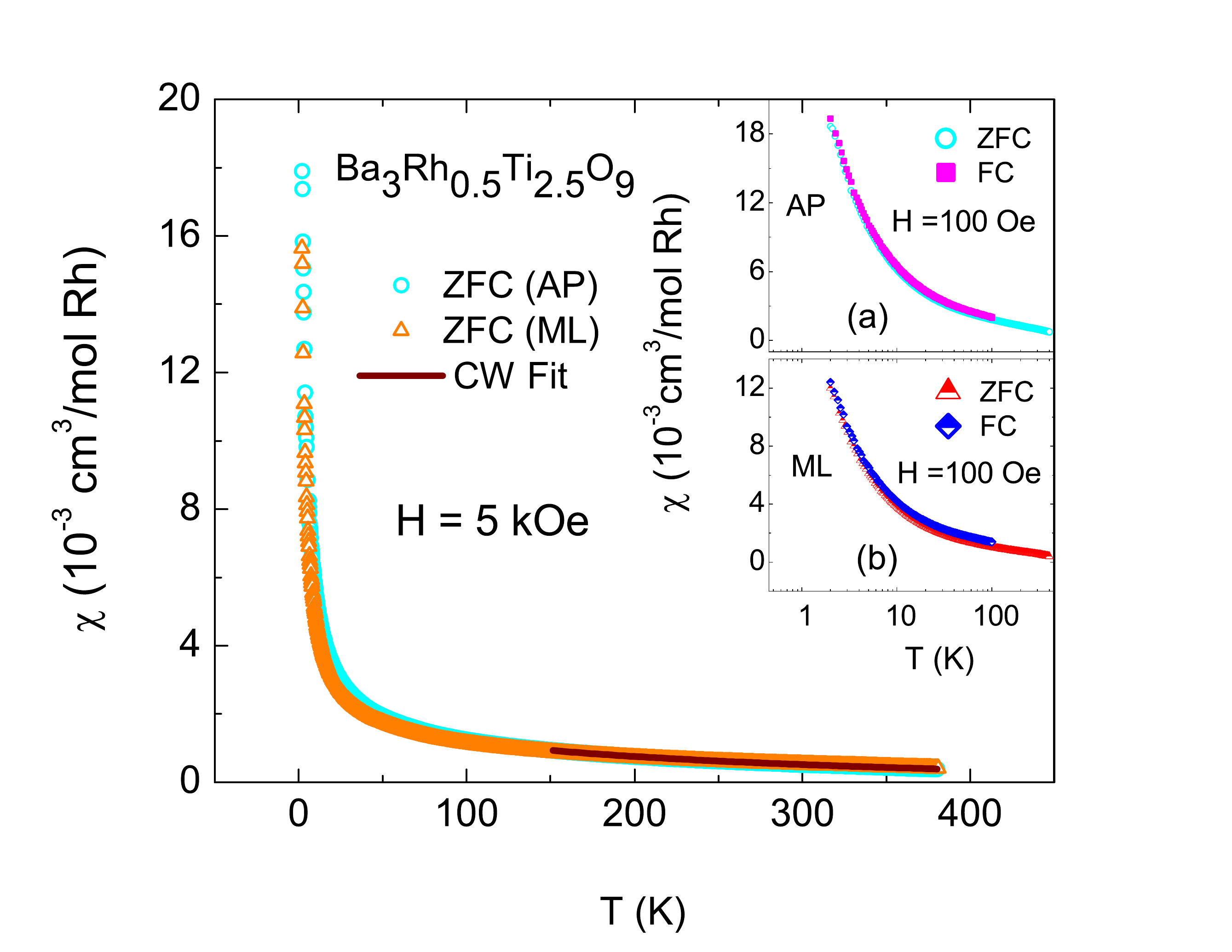}
\par\end{centering}

\protect\caption{(Color online) \label{Rh0.5-SVSM} The main figure shows the temperature
dependent susceptibilities of nominal Ba$_{3}$Rh$_{0.5}$Ti$_{2.5}$O$_{9}$
(AP) (cyan, open circles) and a month older batch of nominal Ba$_{3}$Rh$_{0.5}$Ti$_{2.5}$O$_{9}$
(ML) (orange, open triangles) measured in 5 kOe in the temperature
range $2-400$ K and the brown solid line represents the Curie-Weiss
fit in the temperature range $150-400$ K. Inset (a) and (b) depict
the ZFC and FC data of nominal Ba$_{3}$Rh$_{0.5}$Ti$_{2.5}$O$_{9}$
(AP) and nominal Ba$_{3}$Rh$_{0.5}$Ti$_{2.5}$O$_{9}$ (ML) measured
at $100$ Oe, respectively. }
\end{figure}
The \textit{$\chi(T)$ }of both the as-prepared batch (AP) of nominal
Ba$_{3}$Rh$_{0.5}$Ti$_{2.5}$O$_{9}$ and few months older batch
(ML), exposed to air, of nominal Ba$_{3}$Rh$_{0.5}$Ti$_{2.5}$O$_{9}$
did not show any difference in the ZFC-FC data (see the insets (a)
and (b) of Fig.~\ref{Rh0.5-SVSM}), respectively. The possibility
of sample evolution/degradation with time/exposure to atmosphere was
examined as this is known to happen in the high-Tc cuprates, the Na$_{x}$CoO$_{2}$
system \cite{28-NaxCo2}, as also in the iridate material Na$_{4}$Ir$_{3}$O$_{8}$
\cite{29-Na4Ir3O8 metallic}. Susceptibility data analysis for Ba$_{3}$Rh$_{0.5}$Ti$_{2.5}$O$_{9}$
as prepared (AP) batch gives $\chi_{0}=-3.295\times10^{-4}$ cm$^{3}$/mole
Rh, $\mu_{eff}\sim1.78\pm0.18\,\mu_{B}$, and $\theta_{CW}\sim-154$~K.
The $\chi(T)$ data for Ba$_{3}$Rh$_{0.5}$Ti$_{2.5}$O$_{9}$ and
Ba$_{3}$Ir$_{0.5}$Ti$_{2.5}$O$_{9}$ suggest the formation of a
quantum spin-liquid-like ground state in them. Also, neutron diffraction
measurements done at 2~K, well below $\theta_{CW}$, do not show
any signatures of magnetic order for both the Rh-based compounds,
and are in line with the susceptibility measurements. A comparison
of $\mu_{eff}$ and $\theta_{CW}$ of the various Ir-based and Rh-based
samples of this study are summarized in Table~\textcolor{black}{\ref{Compared Curie}.}

\subsubsection{Heat capacity}

\begin{figure}
\begin{centering}
\includegraphics[scale=0.355]{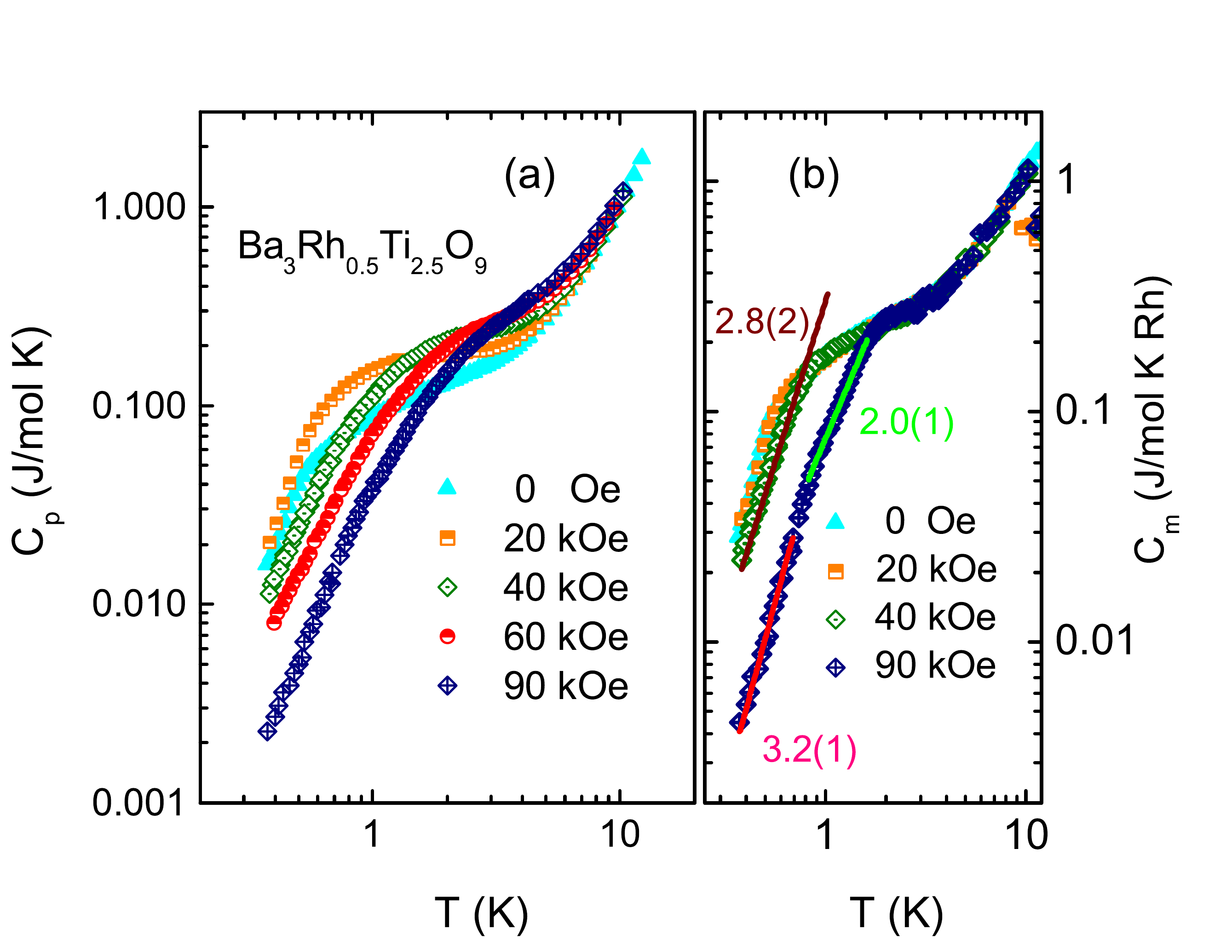}
\par\end{centering}

\protect\caption{\label{Fig Rh0.5-Cp(T)} (Color online). (a) The heat capacity data
for nominal Ba$_{3}$Rh$_{0.5}$Ti$_{2.5}$O$_{9}$ (AP/Baked) in
log-log plot as a function of temperature at different fields. (b)
The magnetic heat capacity $C_{m}$ is shown in a log-log plot and
the solid lines are fits to the power law: $C_{m}=\gamma T^{\alpha}$.
The obtained exponent values are written against the $C_{m}$ data.}
\end{figure}

\begin{figure}
\begin{centering}
\includegraphics[scale=0.355]{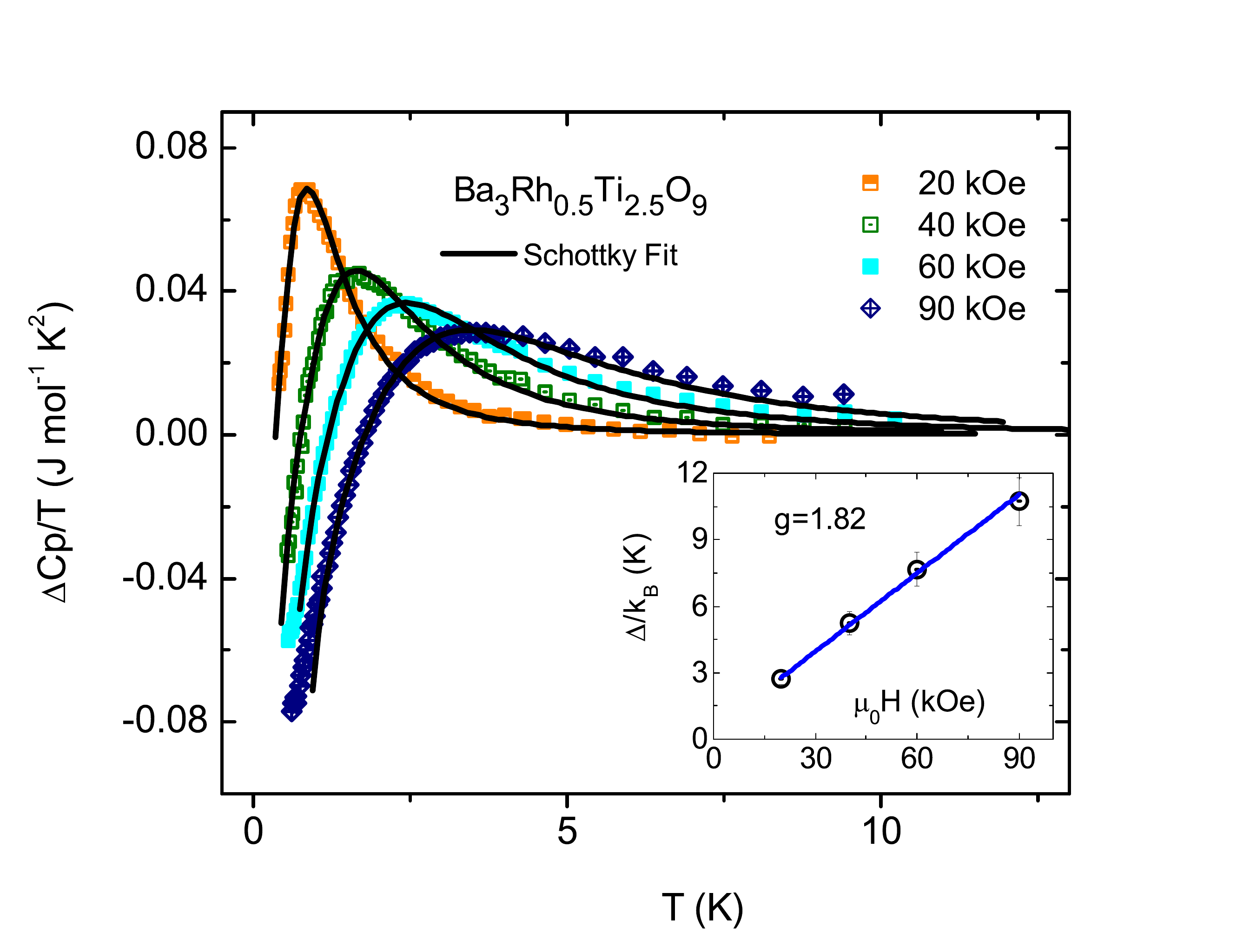}
\par\end{centering}

\protect\caption{\label{Fig Rh0.5 Schottky} (Color online). A few representative plots
of $[C_{P}(H)-C_{P}(H=0)]/T$ or ($\triangle C_{P}/T$) for nominal
Ba$_{3}$Rh$_{0.5}$Ti$_{2.5}$O$_{9}$ at various values of the field
$H$ (as given in the legend) and their fit to Eq.~\ref{eq:1} (black
lines). Inset: the variation of the Schottky gap as a function of
magnetic field for orphan spins. }
\end{figure}

Now we present the$C_{p}(T)$ data obtained for nominal Ba$_{3}$Rh$_{0.5}$Ti$_{2.5}$O$_{9}$
which is single phase. The sample was first subject to heating at
$200$$^{\circ}$C for $10$ hours. After the sample cooled back to
room temperature, $C_{p}(T)$ measurements were performed soon after.
To emphasize the low temperature $C_{p}(T)$ data of nominal Ba$_{3}$Rh$_{0.5}$Ti$_{2.5}$O$_{9}$,
the data are only shown in the $T$-range $0.35-10$~K, as can be
seen in Fig.~\ref{Fig Rh0.5-Cp(T)}(a). The data are qualitatively
similar to those for Ba$_{3}$Ir$_{0.5}$Ti$_{2.5}$O$_{9}$. In order
to extract the magnetic heat capacity $C_{m}$, a similar process
as mentioned before for nominal Ba$_{3}$Ir$_{0.5}$Ti$_{2.5}$O$_{9}$
was adopted. The estimation of Schottky contribution for nominal Ba$_{3}$Rh$_{0.5}$Ti$_{2.5}$O$_{9}$
from experimental data is shown in Fig.~\ref{Fig Rh0.5 Schottky}.
The extracted orphan spin contribution ($f$) for Ba$_{3}$Rh$_{0.5}$Ti$_{2.5}$O$_{9}$
(nominal) amounts to $3$\%-$4$\% of $S=1/2$ entities. The $g$-factor
estimated from plotting the Schottky gap data as a function of field
for nominal Ba$_{3}$Rh$_{0.5}$Ti$_{2.5}$O$_{9}$ (see inset of
Fig.~\ref{Fig Rh0.5 Schottky}) yields $\sim$$1.82$. A value close
to $2$ suggests that the orphan spins do not suffer the effect of
spin-orbit coupling. 

\begin{figure}
\begin{centering}
\includegraphics[scale=0.37]{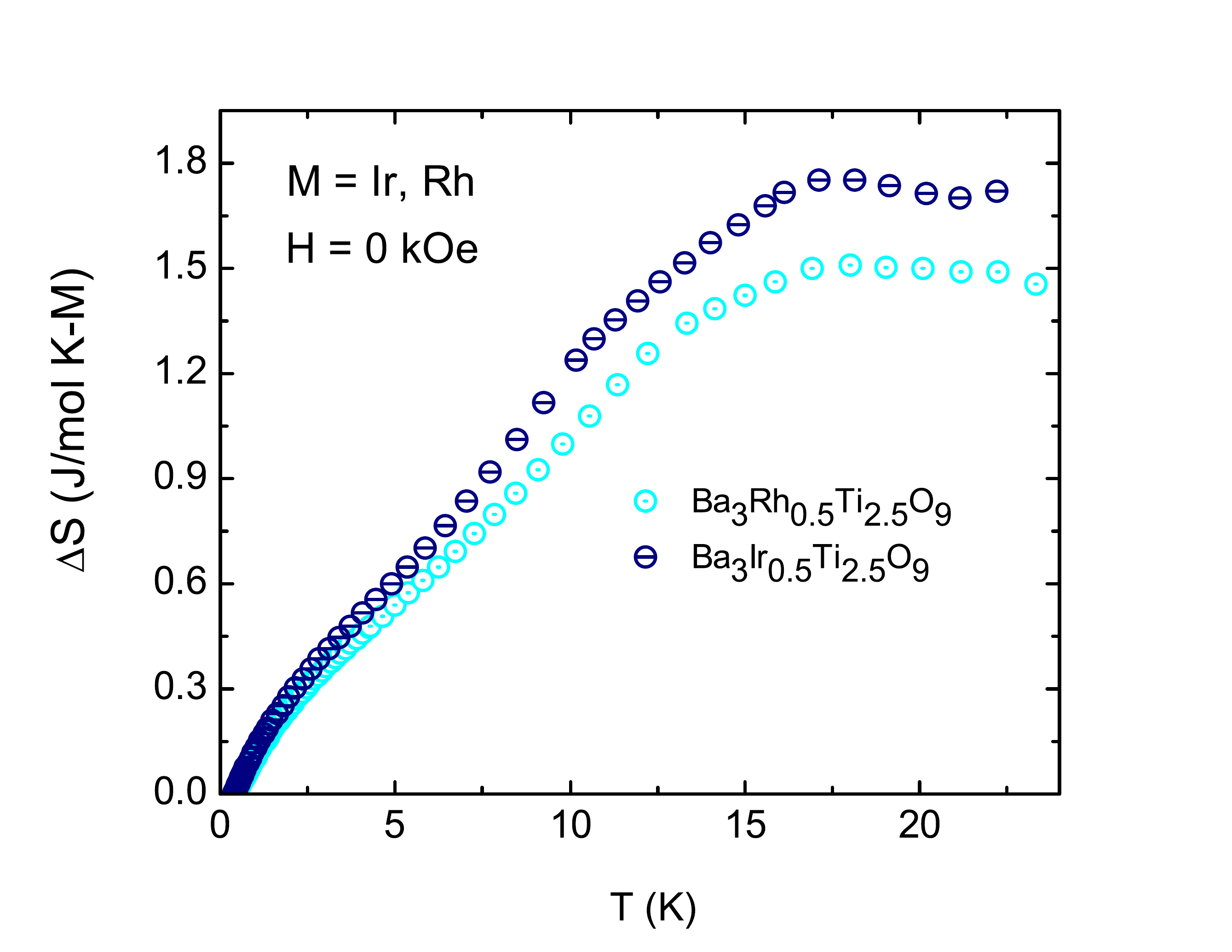}
\par\end{centering}

\protect\caption{\label{Entropy change} (Color online). Entropy change as a function
of temperature for Ir and Rh variants at zero magnetic field for the
nominal compositions of Ba$_{3}$Ir$_{0.5}$Ti$_{2.5}$O$_{9}$ and
Ba$_{3}$Rh$_{0.5}$Ti$_{2.5}$O$_{9}$.}
\end{figure}

The resulting $C_{m}$ data for nominal Ba$_{3}$Rh$_{0.5}$Ti$_{2.5}$O$_{9}$
as a function of $T$ and at various $H$ are shown in Fig.~\ref{Fig Rh0.5-Cp(T)}(b).
Here again the $C_{m}$ for nominal Ba$_{3}$Rh$_{0.5}$Ti$_{2.5}$O$_{9}$
exhibits a variation similar to that of nominal Ba$_{3}$Ir$_{0.5}$Ti$_{2.5}$O$_{9}$
and can be fit with a power law in the low-$T$ range. \textcolor{black}{The
exponent in the power-law at low temperature is somewhat $H$-dependent.}
Furthermore, the estimated entropy change for nominal Ba$_{3}$Rh$_{0.5}$Ti$_{2.5}$O$_{9}$
(see Fig.~\ref{Entropy change}) does not recover the full value
$5.76$ J/mol K expected for ordered spin-1/2 moments. This downshift
in entropy is a common tendency of frustrated spin systems. Similarly,
below $20$~K, the estimated entropy change for Ba$_{3}$Rh$_{0.5}$Ti$_{2.5}$O$_{9}$
comes out to be $1.78$ J/mol K Rh. A comparison of entropy change
for  the studied compounds is tabulated in Table\textcolor{black}{~\ref{Compared Curie}}.

\begin{figure}
\begin{centering}
\includegraphics[scale=0.35]{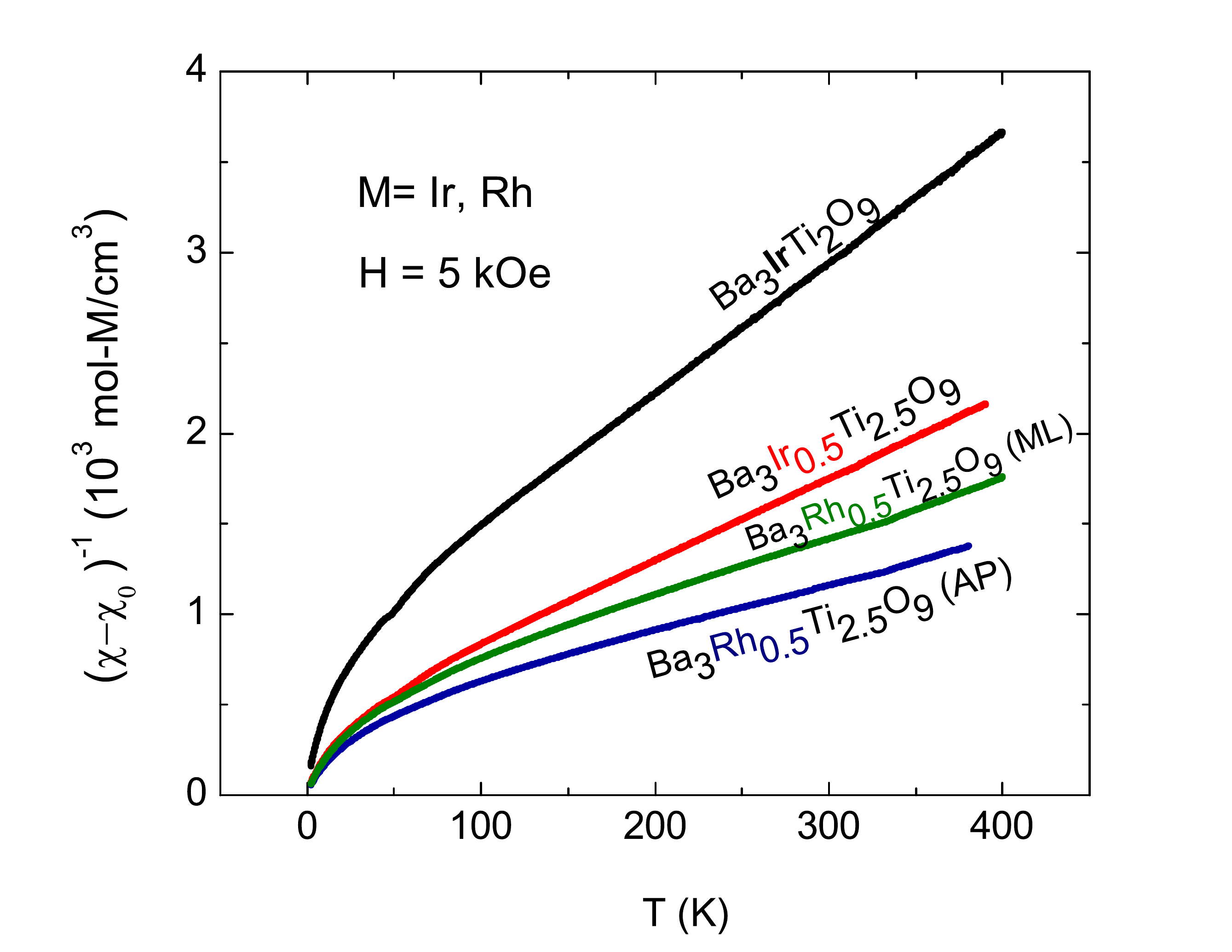}
\par\end{centering}

\protect\caption{(Color online).\label{Inver Chi and Entropy} Inverse susceptibility
plots for Ir, Rh and their variant samples (nominal composition) as
a function of temperature. The data for Ba$_{3}$IrTi$_{2}$O$_{9}$
are taken from Ref. \cite{4-Ba3IrTi2O9}. }
\end{figure}

\section{Discussion}

\begin{table*}
\protect\caption{\label{Compared Curie} Estimated values of the Curie-Weiss temperature,
the effective magnetic moment, the entropy change, and the magnetic
heat capacity exponent for various compounds studied in this paper. }

\centering{}%
\begin{tabular}{|>{\raggedright}p{2.5cm}|>{\raggedleft}p{1.6cm}|>{\raggedright}p{1.7cm}|>{\raggedright}p{1.6cm}|>{\raggedright}p{1.6cm}|>{\raggedright}p{1.6cm}|}
\hline 
Compound & $\theta_{CW}$($K$) & $\mu_{eff}$($\mu_{B}$) & $\Delta S$

(J/mol-K) & $C_{m}(T)$

exponent $\alpha$ & Reference\tabularnewline
\hline 
Ba$_{3}$IrTi$_{2}$O$_{9}$ & $-107$  & 1.05$\pm$0.04 & 0.57 & 1.9 & \cite{4-Ba3IrTi2O9}\tabularnewline
\hline 
Ba$_{3}$Ir$_{0.5}$Ti$_{2}$O$_{9}$ & $-125$  & 1.32$\pm$0.05 & 1.76 & 2.5 & This work\tabularnewline
\hline 
Ba$_{3}$Rh$_{0.5}$Ti$_{2.5}$O$_{9}$

(AP) & $-162$  & 1.78$\pm0.18$ & 1.51 & 3.2 & This work\tabularnewline
\hline 
Ba$_{3}$Rh$_{0.5}$Ti$_{2.5}$O$_{9}$ (ML) & $-149$  & 1.58$\pm0.16$ & not measured &  & This work\tabularnewline
\hline 
\end{tabular}
\end{table*}

\textcolor{black}{There has been a significant amount of work on a
triangular variety of antiferromagnetic systems in the last few years.
One issue that has been persistent among these systems, such as Ba$_{3}$CuSb$_{2}$O$_{9}$
(triangle or honeycomb) \cite{13- Ba3CuSb2O9,18 Ba3Cu Nakasduji},
Herbertsmithite (Kagome) \cite{30- Herbert,31 Mend JPSJ,32 Olariu NMR},
SrCr$_{9x}$Ga$_{12-9x}$O$_{19}$ (Kagome) \cite{33 SCGO Ga NMR,34 Broholm SCGO},
Ba$_{3}$IrTi$_{2}$O$_{9}$ (triangular biplanes) }\cite{4-Ba3IrTi2O9}\textcolor{black}{,
or Sc$_{2}$Ga$_{2}$CuO$_{7}$ \cite{35 SGCO IITB,36-NMR SGCO} (triangular
biplanes), is the effect of vacancies at the magnetic sites. In minute
amounts, these might simply give rise to an additional Curie term
in the susceptibility and a Schottky like contribution to the heat
capacity, keeping the correlated ground state intact. Experimental
studies on the effect of percolation in magnetic systems are often
limited to concentrations of dopants (nonmagnetic ions) where the
LRO persists. In this region, the ordering temperature first falls
linearly with an increase in the dopant concentration. Below the percolation
threshold, magnetic LRO is destroyed. What might happen for even larger
defect concentrations? It appears that there has been no experimental
report on the work on (frustrated) triangular systems near the percolation
threshold. A study on the layered honeycomb system Mn$_{x}$Zn$_{1-x}$PS$_{3}$
\cite{37-percolation} finds that below the percolation threshold,
the high-temperature susceptibility data can be modeled as a combination
of a term due to a randomly diluted antiferromagnet and a Curie term
from uncompensated spins. The effective Curie-Weiss temperature and
the Curie constant from the first term were found to }\textit{\textcolor{black}{decrease}}\textcolor{black}{{}
with increasing dopant concentration. In a frustrated systems such
as ours, one might expect to have a frozen/spin-glass state for a
small concentration of nonmagnetic ions (i.e., above the percolation
threshold) in case ordering is prevented. On going below the percolation
threshold, a scenario as in Mn$_{x}$Zn$_{1-x}$PS$_{3}$ might present
itself in case of random occupation of Ir and Ti (in our system).
For a bi-layer triangular system (our case), the percolation threshold
might be somewhat lower than that of a triangular lattice (site percolation
p$_{c}$= 0.5). In any case, for the two Ir (or Ti) concentrations
studied by us (magnetic ion fraction 1/2 and 1/4) the Curie-Weiss
temperature and the Curie constant }\textit{\textcolor{black}{increase}}\textcolor{black}{{}
slightly with increasing Ti concentration. For the Ba$_{3}$Rh$_{0.5}$Ti$_{2.5}$O$_{9}$
composition, a similar Curie-Weiss temperature was obtained. In case
of a statistical occupancy, the probability of finding $r$ Ir neighbours
in the plane for an Ir (or Rh) within a plane is $^{6}C_{r}(p)^{r}(1-p)^{6-r}$
where $p$ is the fraction of Ir (or Rh).}\textcolor{red}{{} }\textcolor{black}{Clearly,
the probabilities of finding various clusters change significantly
when $p$ changes from 1/2 to 1/4. However, neither our Curie constants
(or Curie-Weiss temperatures) nor the fraction contributing to the
Schottky anomaly in the heat capacity change significantly. This suggests
that Ti and Ir (or Rh) occupancy in the biplanes might not be at random.}\textcolor{red}{{}
}\textcolor{black}{In fact, in Ba$_{3}$RuTi$_{2}$O$_{9}$, electronic
structure calculations \cite{38-Ru-Ti} have suggested that Ru-Ti
dimers are energetically more stable than Ru-Ru or Ti-Ti dimers. A
similar argument may be applicable for Ba$_{3}$IrTi$_{2}$O$_{9}$
(as also Ba$_{3}$Ir$_{0.5}$Ti$_{2.5}$O$_{9}$ and Ba$_{3}$Rh$_{0.5}$Ti$_{2.5}$O$_{9}$)
and there might be some site ordering within the biplanes which could
lead to continued connectivity in spite of the apparent depletion.}\textcolor{blue}{{}
}\textcolor{black}{Further work such as determination of the pair
distribution function via high statistics diffraction data would be
useful to settle this matter. Another possibility arises from the
recent effort to explore the role of disorder in stabilising a spin
liquid state in systems such as the subject of these papers (\cite{19-MILA,39-AFM melt}).
These have suggested that orphan spins that result from vacancy/disorder
might remain delocalised and hence actually help in preventing freezing
and stabilise a spin liquid state. Another point to note is that the
effective moments (estimated from the Curie terms) are much smaller
than those expected from pure spin-only values. This is possibly due
to a spin-orbit coupling since it is seen that in going from 5$d$-based
(Ir) to 4$d$-based (Rh) to 3$d$-based systems (Cu, in SGCO \cite{35 SGCO IITB})
the effective moment goes up with the full spin only value being attained
for the 3$d$ case. In all the above situations, the spin-liquid state
seems to be preserved suggesting that it is the lattice geometry which
drives this rather than any spin-orbit coupling effect. Electron or
hole doping might be interesting to tune the ground states of these
systems.}

\section{Conclusions}

In summary, \textcolor{black}{we have introduced a new 4}\textit{\textcolor{black}{d}}\textcolor{black}{{}
based, Rh$^{4+}$ ($S=1/2$), bi-triangular lattice system} Ba$_{3}$Rh$_{0.5}$Ti$_{2.5}$O$_{9}$
(nominal) and in addition, also have studied the effect of dilution
on its 5\textit{d} analogue Ba$_{3}$Ir$_{0.5}$Ti$_{2.5}$O$_{9}$
(nominal) via x-ray, neutron diffraction, magnetization, and heat
capacity measurements. Our xrd/neutron refinement results infer the
presence of highly depleted magnetic biplanes. The \textit{dc} susceptibility
measurements do not show any signature of magnetic order, glassiness
or spin-freezing behavior for the freshly prepared/baked samples.
Further, the measured heat capacity down to $0.35$ K, well below
the exchange energy, is found to be free from any kind of magnetic
order and suggests the persistence of a QSL ground state for all of
the materials studied. The novelty of these 4\textit{d}/5\textit{d}
based materials is that in spite of the large depletion, they maintain
strong antiferromagnetic couplings and retain the QSL ground state.
Further local probe measurements like NMR and $\mu SR$ will be of
great importance to shed deeper insight into the intrinsic susceptibility
(in presence of disorder) and the homogeneity of the ground state,
respectively.

\section{acknowledgement}

\textcolor{black}{We thank Department of Science and Technology, Govt.
of India and the Indo-Swiss joint research programme, the Swiss National
Science Foundation and its SINERGIA network MPBH for financial support.}\textcolor{blue}{{}
}R. Kumar acknowledges CSIR (India) and IRCC (IIT Bombay) for awarding
him research fellowships to carry out this research work. This work
is partly based on experiments performed at the Swiss spallation neutron
source SINQ, Paul Scherrer Institut, Villigen, Switzerland.

\end{document}